\documentclass[]{amsart}
\usepackage{latexsym,fancyhdr,amssymb,color,amsmath,amsthm,graphicx,listings,comment}
\usepackage[section]{placeins}
\let\paragraph\subsection

\title{Shannon capacity, Chess, DNA and Umbrellas}
\author{Oliver Knill}
\date{August 10, 2021}
\address{Department of Mathematics \\ Harvard University \\ Cambridge, MA, 02138 }
\subjclass{05CXX,  % Graph theory 
           94-XX,  % Information and communication theory
           68R10}  % Graph theory in computer science

\begin{document}
\maketitle

\begin{abstract}
A vexing open problem in information theory is to find the Shannon capacity 
of odd cyclic graphs larger than the pentagon and especially for the heptagon. 
Lower bounds for the capacity are obtained by solving King chess puzzles.
Upper bounds are obtained by solving entanglement problems, that is to 
find good Lovasz umbrellas, quantum state realizations of the graph. We observe
that optimal states are always pure states. The rest is expository. 
One general interesting question is whether the Shannon capacity 
is always some n-th root of the independence number of the n'th power of the graph.
\end{abstract}

\section{In a nutshell} 

\begin{figure}[!htpb]
\parbox{11cm}{
\parbox{5.5cm}{\scalebox{0.04}{\includegraphics{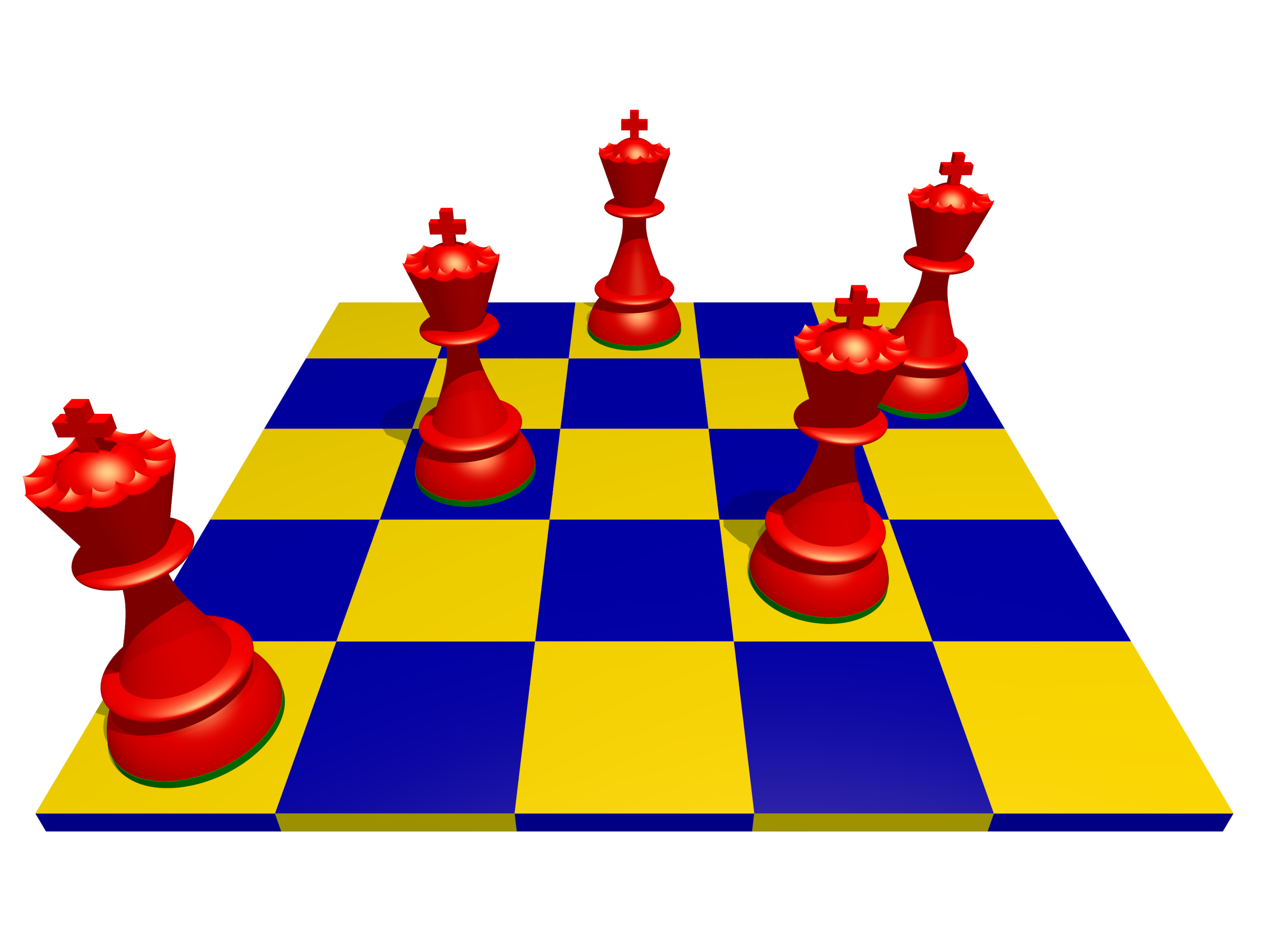}}}
\parbox{5.5cm}{\scalebox{0.6}{\includegraphics{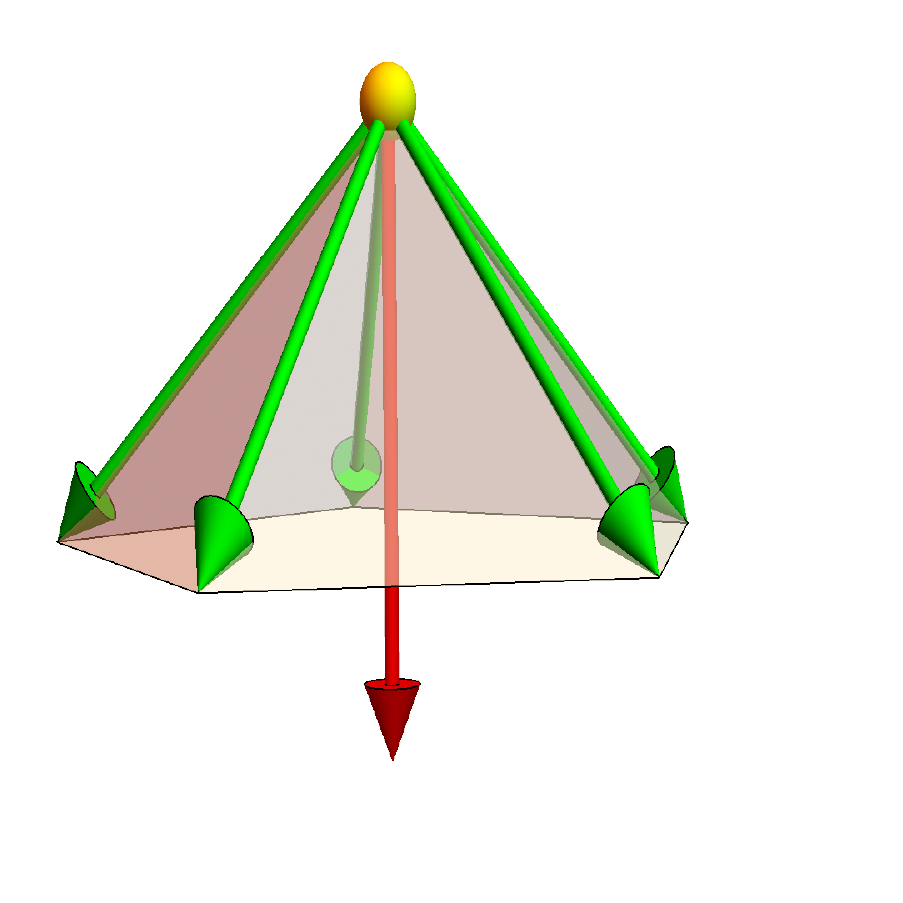}}}
}
\label{King}
\caption{
This picture illustrates the proof that $\Theta(C_5)=\sqrt{5}$. The $5$-king configuration
gives the lower bound $\sqrt{5}$, a Lovasz umbrella gives the upper bound $\sqrt{5}$. 
To determine the unknown $\Theta(C_7)$ we need to find king configurations on a toroidal
$\mathbb{Z}_7^d$ board which has a matching umbrella on some space $\mathbb{S}^q$ of 
quantum states. We can only wonder whether $\Theta(C_7)=\alpha(C_7^p)^{1/p}$ 
for some $p$ or whether $\Theta(C_7)$ is equal to the Lovasz number 
$\theta(C_7)=7 \cos(\pi/7)/(1+\cos(\pi/7))=3.31767...$ of $C_7$. 
% n=7; n Cos[Pi/n]/(1+Cos[Pi/n]) 
}
\end{figure}

\paragraph{}
The {\bf Shannon capacity} $\Theta(G)=\lim_{n \to \infty} \alpha(G^n)^{1/n}$ \cite{Shannon1956} measures
the exponential growth rate of the {\bf independence number} $\alpha(G^n)$ of {\bf Shannon powers} $G^n$ of a finite
simple graph $G=(V,E)$. The power $G^n$ is the graph with the Cartesian product $V^n$ as vertices and where two points
are connected if all of the projections on the individual coordinates are connected or identical. For example, 
if $G$ is a path graph with $8$ nodes, then $G^2$ is a standard {\bf king chess graph}, where a vertex in the form of 
a square of the standard $8 \times 8$ chess board is connected to all king-neighboring squares. 
The Shannon capacity is already unknown
for odd circular graphs $C_p$ larger than the pentagon $C_5$. When constructing robust {\bf block codes},
the graph takes the letters $V$ of an alphabet as nodes and builds connections $E$ of letters
if they have a danger to be confused. In order to find lower bounds for the capacity, we need to 
compute or estimate the {\bf independence number} $\alpha(C_p^n)$ of the n'th power, which is the maximal number of 
{\bf non-interacting kings} that can be placed on a $n$-dimensional 
{\bf toroidal chess} board $C_p^n$. 

\paragraph{}	
In order to find upper bounds for the capacity, Shannon already used $\alpha(G) \leq \rho(G)$,
where $\rho(G)$ is the {\bf Rosenfeld number} \cite{Rosenfeld1967}, 
the minimal $\sum_{v \in V} f(v)$, where $f$ ranges over all non-negative functions
with $\sum_{v \subset x} f(v) \leq 1$ for all cliques \cite{Hales1973}; this itself is bounded above by $\sigma(G)$, the 
{\bf clique covering number}. Better even is to construct {\bf Lovasz umbrellas} $U:V(G) \to H$ \cite{Lovasz1979}
which assigns a {\bf quantum state} $U(x)$, a unit vector in a Hilbert space $H$ of {\bf density matrices},
to each node $x$ and also fixes a {\bf vacuum state} $c$, the {\bf umbrella stick}. The condition to be satisfied
is the {\bf de-correlation property} $(U(x) \cdot U(y))=0$ for all $(x,y) \notin E$. The Lovasz number $\theta(G)$ is 
also interesting in graph coloring as the clique number $\omega(G)$, the chromatic number $X(G)$ satisfy
$\omega(G) \leq \theta(\overline{G}) \leq X(G)$ which is known as the {\bf Lovasz Sandwich theorem}. It is equivalent to
$\alpha(G) \leq \theta(G) \leq \sigma(G)$, where $\sigma(G)$ is the {\bf clique covering number}. 

\paragraph{}
A {\bf quantum mechanical interpretation} of the Lovasz umbrella is to think of a node as a point in space
and $U(x)$ as the {\bf quantum state} attached of this point. The orthogonality condition is that 
quantum mechanical states assigned to non-adjacent vertices $x,y$ must be uncorrelated. 
It is a natural {\bf causality condition}. When multiplying graphs $G \cdot H$ using the {\bf strong product}
(we also call it Shannon product, as Shannon was the first to define it in 1956. Complementary 
is the {\bf large product} which we also call Sabidussi product because of \cite{Sabidussi} from 1959),
the states {\bf tensor multiply} $U \otimes V(x,y) = U(x) V(y)$, the 
orthogonality condition is inherited. More generally, if $U(x)$ are {\bf density matrices} (self adjoint, positive semi-definit
trace-class operators of trace $1$),
encoding possibly {\bf entangled states}, then the density matrix of the graph product is the tensor product 
of the density matrices and the dot product $(U(x) \cdot U(y)) = {\rm tr}(U(x)^T U(y))$ is a covariance.
The {\bf Lovasz number} $\theta(G) = \inf_{U,c} \max_{x} (c \cdot U(x))^{-2}$ therefore is
{\bf multiplicative} $\theta(G H) = \theta(G) \theta(H)$. It is 
an upper bound for any $\alpha(G^n)^{1/n}$ counting the maximal number of an independent set
$I \subset V(G)$ of the graph because the Parcsval's inequality $\sum_{x \in I} (c \cdot U(x))^2 \leq (c \cdot c) = 1$
implies $\theta(U)={\rm min}_x (c \cdot U(x))^2 \leq 1/|I|$ or $|I| \geq {\rm max}_x 1/(c \cdot U(x))^2$. It follows
from the compatibility with product that $\alpha(G^n)^{1/n} \leq \Theta(G) \leq \theta(G)$. 

\paragraph{}
{\bf Density matrices} are self-adjoint real matrices which have non-negative spectrum and which have trace $1$. The
eigenvalues being non-negative and adding up to $1$ can be interpreted as a probability distribution on the
vertex set. The {\bf Lovasz number} $\theta(G)$ quantifies the {\bf minimal correlation} which 
pure or entangled states can have with the vacuum state $c$. 
The two variational tools to bound the Shannon capacity are sharp 
for the pentagon: one can place five non-interacting kings on a $5^2 = 5 \times 5$ torus board and construct an 
explicit Lovasz umbrella $(U,c)$ in $\mathbb{R}^3$ with the same upper bound.
This is how Lovasz showed in \cite{Lovasz1979} that the capacity of the pentagon is $\sqrt{5}$.
Do optimal umbrellas always consist of {\bf pure states} $U(x)$? 

\paragraph{}
For the {\bf heptagon} $C_7$, the Shannon capacity is not known. While the pentagon is self-dual $\overline{C_5}=C_5$,
the {\bf graph complement} $\overline{C_7}$ of the heptagon $C_7$ is the {\bf M\"obius strip}. 
When looking at the computation of the Shannon capacity of the pentagon, a natural {\bf lock-in} conjecture: 
there should be an integer $p$ such that the independence number $\alpha(C_7^p)$ of $C_7^p$ is equal to the capacity. 
One can also ask whether there is an Lovasz umbrella $(U,c)$ for $C_7^p$ 
which has an {\bf umbrella opening number} $\theta_{U,c}(C_7^p)={\rm min}_x (c \cdot U(x))^2$ 
which  matches the Shannon capacity. For general graphs, the Lovasz number is 
not a sharp upper bound. \\

\begin{center}
{\bf Lock-in conjecture:} For every $n$, there exists $p$
such that $\alpha(C_n^p)^{1/p} = \Theta(C_n)$
\end{center} 

\paragraph{}
The question can be asked for any finite simple graph $G=(V,E)$. 
No example seems to be known, where not $\alpha(G^p) = \Theta(G)$ for some $p$.
For the upper bound, \cite{Haemers1978} has shown that the capacity can be strictly 
smaller than the Lovasz bound. Haemers introduced as a new upper bound, the rank 
of a $n \times n$ matrix $B$ that {\bf fits} the graph in the sense that $B_{ij}=0$ if
$i,j$ are adjacent and $B_{ii} \neq 0$. The {\bf Haemers bound} is the minimum of all ranks 
among all such matrices. 

\begin{figure}[!htpb]
\scalebox{0.15}{\includegraphics{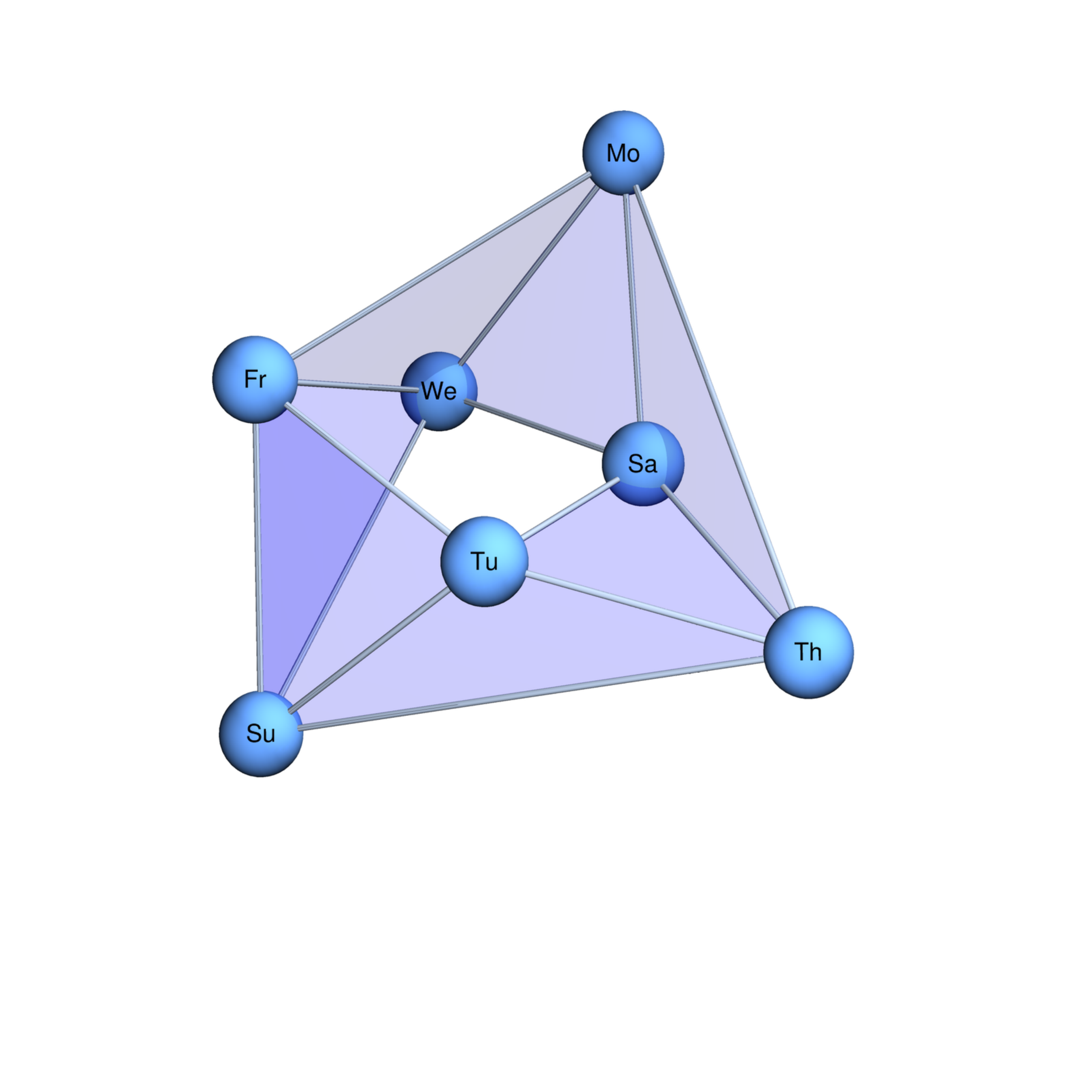}}
\label{DNA}
\caption{
The complement of the $7$-gon is the M\"obius strip. It can be illustrated by assigning 
to each of the 7 vertices a day of the week and connect two days, if they are not adjacent. 
Understanding the Shannon capacity could come also from knowing the clique number growth of 
Sabidussi powers $G^n$ of this graph. We currently do not know how to compute this even for 
small $n$. 
}
\end{figure}

\paragraph{}
The reason why we do not know the Shannon capacity of $C_7$ is that 
we do not now how many independent kings we can place on a $7^n$ toroidal chess board for larger $n$ 
(the exact value seems even not to be known for $n=3)$. 
We would need an umbrella $G=C_7$ which matches any of the {\bf king packing problems} on 
toroidal chess boards $G^n$. Since independence problems are linked to maximal clique size and
{\bf clique covering problems} are linked to chromatic numbers in the graph complement, it can also help to 
investigate the graph complement of cyclic graphs and {\bf Sabidussi products} of them which correspond to 
Shannon products in the complement. Already the graph complement $\overline{C_n}$
of cyclic graphs $C_n$ are topologically extremely rich:
they are all homotopy spheres $S^m$ or $S^m \wedge S^m$ with $m=(n/3-1)$
where $(x)$ is the closest integer to $x$. 
For some recent work giving the asymptotic of $\Theta(C_p)$ for large $p$ see \cite{Bohman2003}.
It shows that $\Theta(C_{2n+1}) -(n+0.5) \to 0$. So, asymptotically, the capacity of the 
odd cyclic graphs becomes asymptotic to the even ones. 

\section{A genetic code}

\paragraph{}
In order to illustrate how capacity can appear in an application, we look at a hypothetical
{\bf genetic code}. An other illustration appears with \cite{Matousek} which deals with 
a communication optimization problem in spy business. (The booklet \cite{Matousek} was where we
learned first about the topic.)
Let us visit a planet in the Andromeda galaxy, where life encodes its genetic information using
nucleodides like here. Every {\bf gene} is a word in these $5$ letters $A,G,C,T,S$. We assume
that $A$ can easily be confused with $G$, $G$ with $C$, $C$ with $T$, $T$ with $S$ and $S$ with $A$. 
Any {\bf code} using a pair like $A,C$ doubled with the independent pair $G,T$ produces
a encoding which looks like the one we have here on earth
with the four nucleotides {\bf Adenine} A, {\bf Guanine} G, {\bf Cytosine} C 
and {\bf Thymine} T. The independent pair $A,C$ can be doubled with the independent pair $G,T$ 
to allow replication. The nucleodide $S$ is not used yet. 

\paragraph{}
Suddenly, a new form of DNA with a {\bf quadruple helix} is observed, where 
{\bf pairs of pairs of nucleotides} are used to encode the genetic information. 
Scientists sequencing such a new genome see $5$ robust combinations $AA,GC,CS,TG,ST$ 
paired with the other robust combination $CC,GS,SG,AT,GA$ (see Figure~(\ref{DNA}).
The new genetic code still allows both for error correction as well as replication.
While it needs twice as many nucleotides, it is possible to store
more information as before. The code has become more effective. 
Is it possible to pack using triples with a {\bf sextuple helix}? 
The answer is no, as we know the Shannon capacity of $C_5$. The code is already optimal for 5 nucleotides. 
Building larger block codes does not improve the communication capacity. 

\begin{figure}[!htpb]
\scalebox{1.3}{\includegraphics{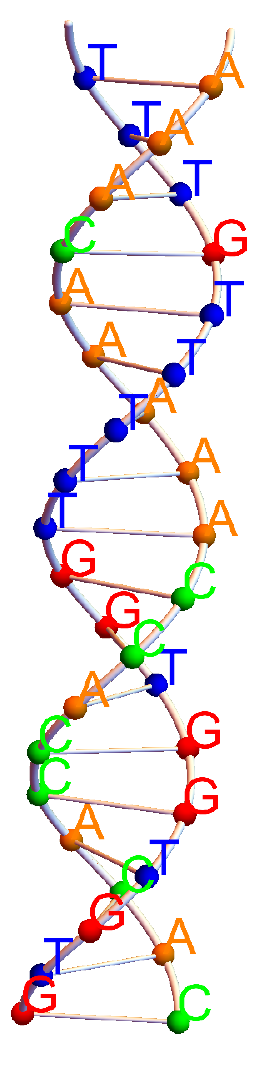}}
\scalebox{1.3}{\includegraphics{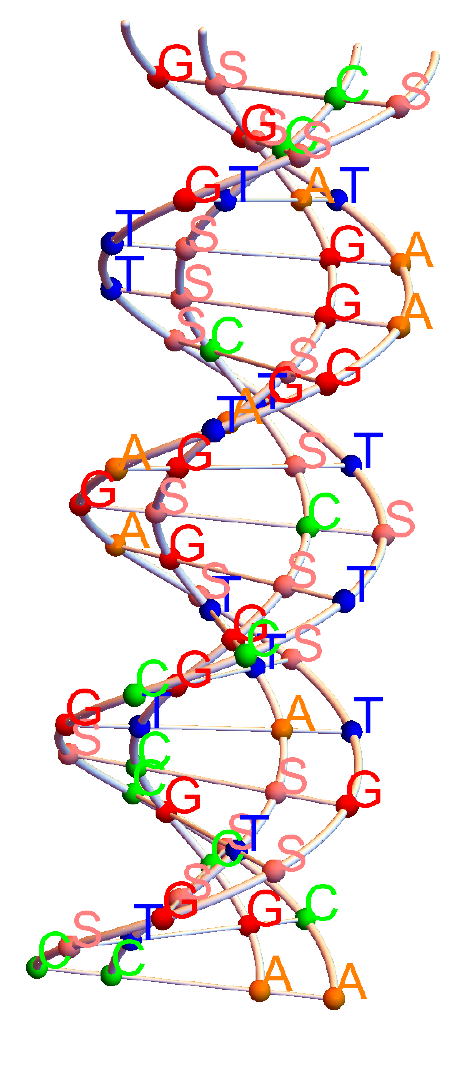}}
\label{DNA}
\caption{
A hypothetical new form of DNA with 5 nucleotides from a galaxy
far, far away. To the left we see the usual encoding which resembles our own DNA code
here on earth.  However, when doubling up the helix, it is possible to store $\sqrt{5}$ times 
more information using only twice as much material. The doubling up is more effective. 
}
\end{figure}

\paragraph{}
Assume we encode information using $7$ proteins $A,B,C,D,E,F,G$
and assume adjacent proteins can be confused: for example, $B$ can 
be mixed up with $A$ and $C$, while $A$ can easily be mixed up with $G$ or $B$  etc.
What is now the optimal encoding? Does it make sense to find a better
block code arrangement? This brings us to chess.

\paragraph{}
As we learned from viewers of the youtube video to this movie,
there are quadruple DNA helix observations also here on earth.
A recent article about this is \cite{Quadruplehelix}. 
There are also bio-molecular {\bf quadruple helix nanowires} that are 
stable in potassium rich conditions \cite{Gquartet}.

\section{Chess variants}

\paragraph{}
The TV mini series ``Queen's Gambit" has sparked new interest
in chess. Chess is a game which combines drama, sport, art and mathematics.
Many variants of chess are known \cite{PritchardChess}. The {\bf Fischer random game} for example
looks at all $960=4*4*6*10$ permutations of the main figures, so that the setup is the same for both players,
so that bishops are on different colors and so that the king is between two rooks. 
Good chess games or chess problems have become pieces of art which similarly as paintings 
are collected and exhibited. 

\paragraph{}
There is also a lot of quite elegant mathematics involved, especially 
if one looks at variants of chess. Chess modifications can change the
dimension, shape or the topology of the board or then the number of initial figures.
One can also change other rules like anti-chess where one wins if one gets his
own king chess-mated. There is no limit in creativity.

\paragraph{}
Independent of the game, one can also study position types.
Many combinatorial problems are chess motivated. Are there closed knight tours
for example or problems like placing $8$ non-interacting queens on a $8 \times 8$ board.
An other interesting problem is to place non-interacting {\bf super queens}, queens which 
additionally also can move like a king. The problem to place as many kings as possible on an 
$7^n=7 \times 7 \times \cdots \times 7$ board is the problem of computing the independence 
number of the king graph.  In the case of a cyclic graph $C_7$, the graph $C_7^n$
has the homotopy type of a $n$ torus. 
The King ``packing question"  is a very {\bf graph theoretical} problem and understanding
this packing problem is required if we want to nail down the open problem of of finding
the Shannon capacity of $C_7$. 

\section{A measure for efficiency}

\paragraph{}
Lets look again at the original Shannon capacity but more from the point of view of chess. 
Shannon observed that multiple encoding can render the channel more
robust. Mathematically, if the alphabet is given by a graph $G$ in which letters which 
can be confused are connected by edges, the problem is to find the independence number
of a product $G \cdot G \cdots \cdot G$ of a graph. 
We can think of it as writing with a new alphabet with $|G|^n$ letters 
identify a set of pairs which can not be confused with each other in the sense that 
we can not change one of the entries to morph one code into an other. 
In the case of $G=C_7$, there is an {\bf independent chess configuration}
with $10$ kings on the board $G^2$. This produces a lower bound $\sqrt{10} =3.16228...$
for $\Theta(C_7)$. We can easily put $30$ kings on a three dimensional torus $G^3$ by
placing $10$ kings in three different floors $1,3,5$ of the 3D chess board. 

\begin{figure}[!htpb]
\scalebox{0.7}{\includegraphics{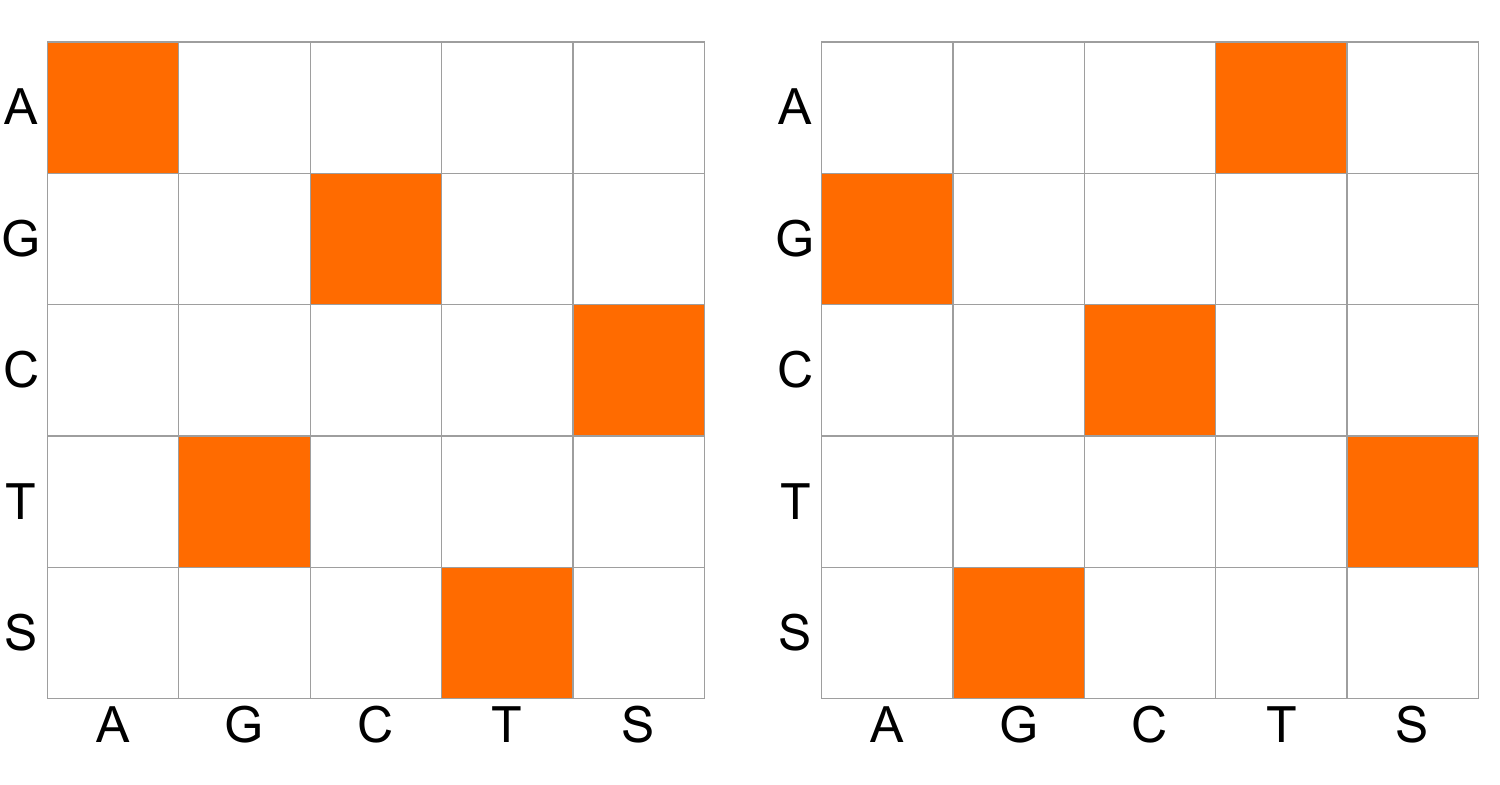}}
\label{DNA}
\caption{
One can pack $\alpha=5$ kings onto a toroidal $5 \times 5$
chess board. The information capacity is $\sqrt{5}$.
It turns out that on a $5^n = 5 \times 5 \times \cdots \times 5$
chess board, one can maximally place $5^{n/2}$ kings. Nobody
knows what happens on a $7^n$ toroidal chess board. Knowing the answer
would give us the Shannon capacity of $C_7$. 
}
\end{figure}

\begin{figure}[!htpb]
\scalebox{0.09}{\includegraphics{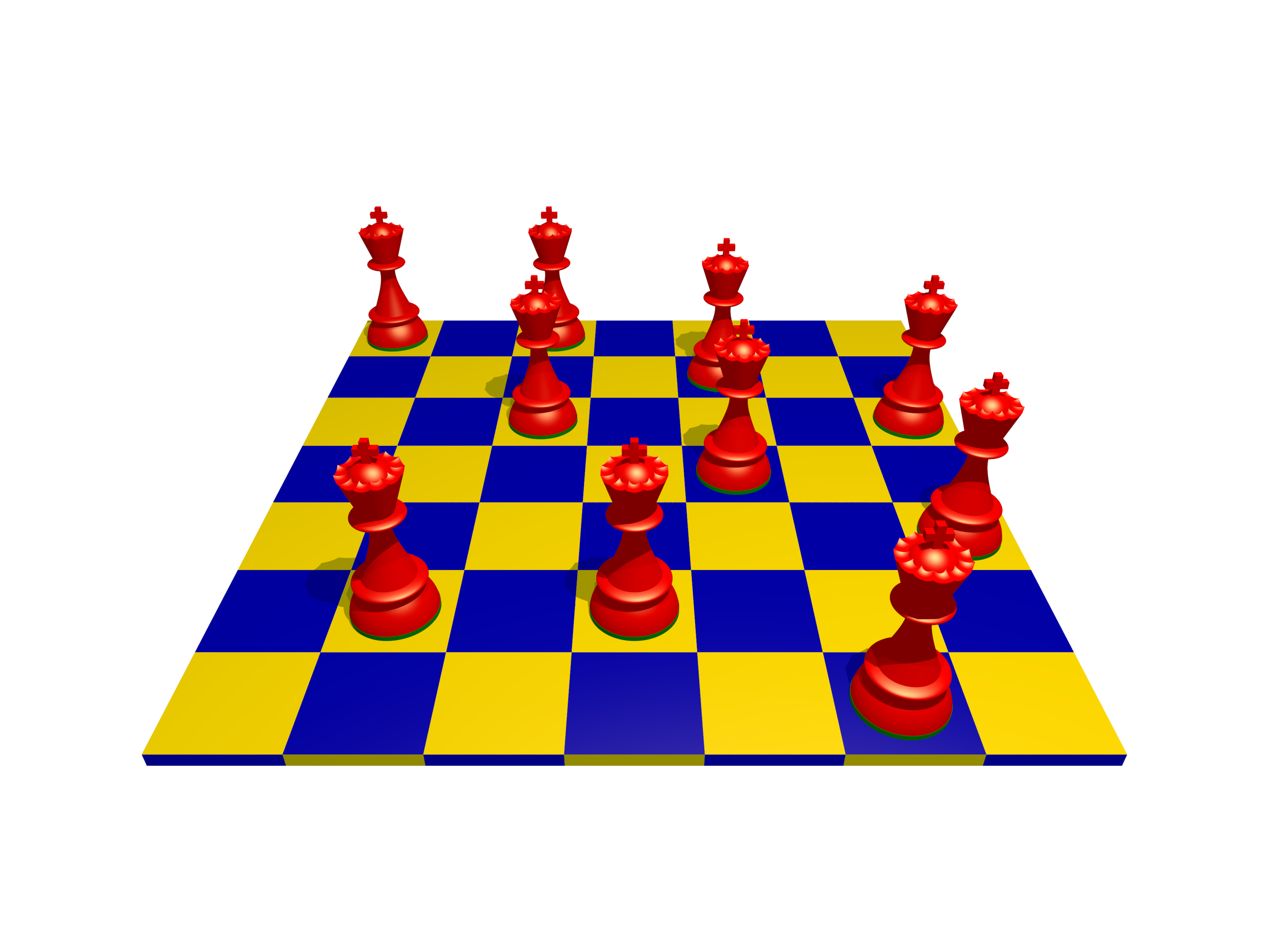}}
\label{7king}
\caption{
One can pack 10 kings in a $7 \times 7$ toroidal chess board. 
This gives a lower bound for the Shannon capacity of $C_7$. 
}
\end{figure}

\paragraph{}
Claude Shannon was not only a brilliant thinker but also skillful tinkerer. He also wrote the first paper 
about building a chess computer and himself built a chess machine \cite{MindAtPlay}. 
Shannon already knew what happens with cyclic alphabets of even length. 
For example, if we have 4 proteins A,G,C,T and neighbor confusions can happen, then
the independent set $A,C$ is safe. In some sense, this is what nature has chosen to do 
in the encoding of information for life on earth. We pair up the four nucleotides. We can ask now
whether it is possible to make the channel more efficient by using pairs. This is the 
problem of how many kings can be placed on a $4 \times 4$ board so that they can not hit each
other. The answer is clear. There are four. We can not increase the capacity by building
redundant channels. Shannon already saw that this is not the case for 5 x 5. In that
case, one can place 5 kings on a toroidal 5 x 5 chess board. 

\section{Calculating with geometries}

\paragraph{}
One can multiply not only numbers, but also geometries. The product of two lines is a plane.
It was Descartes who first introduced this multiplication. Points are now pairs of real numbers.
Shannon multiplied graphs. The product of two graphs takes the Cartesian product of the
vertex sets and connects two such pairs if the projection of the connection on both sides
is either a vertex or edge. If we multiply two cyclic graph we get a discrete torus. Two 
points are connected, if they can be connected by a king move. 

\begin{figure}[!htpb]
\scalebox{0.6}{\includegraphics{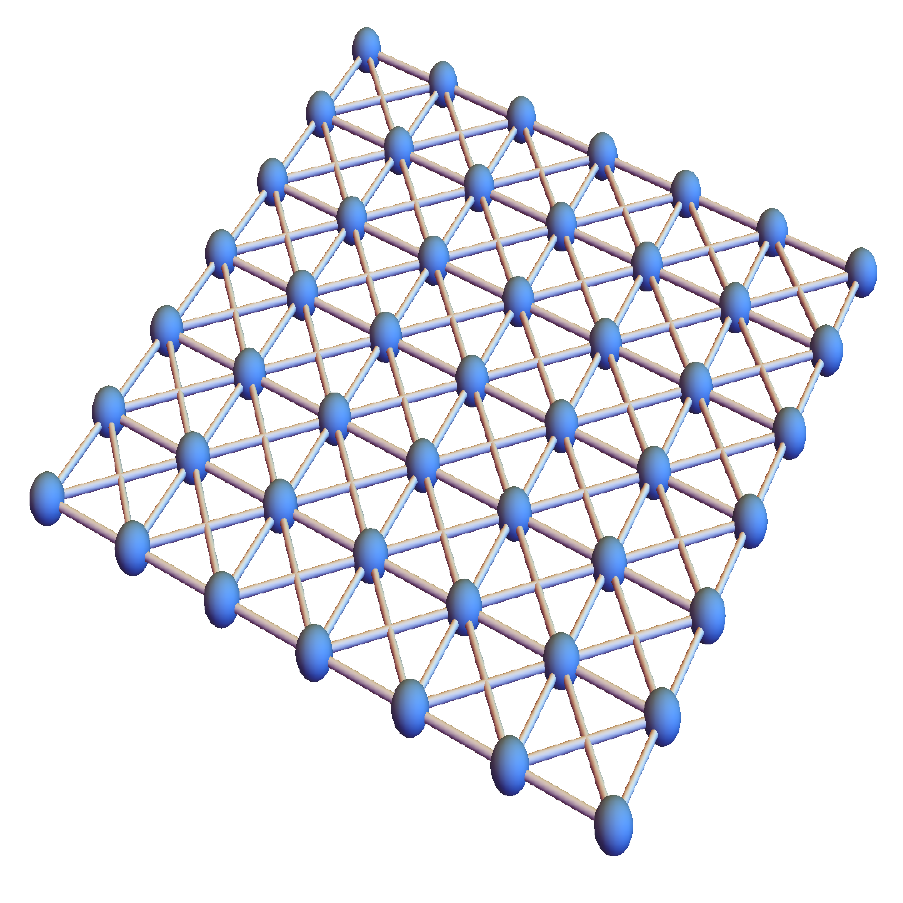}}
\scalebox{0.6}{\includegraphics{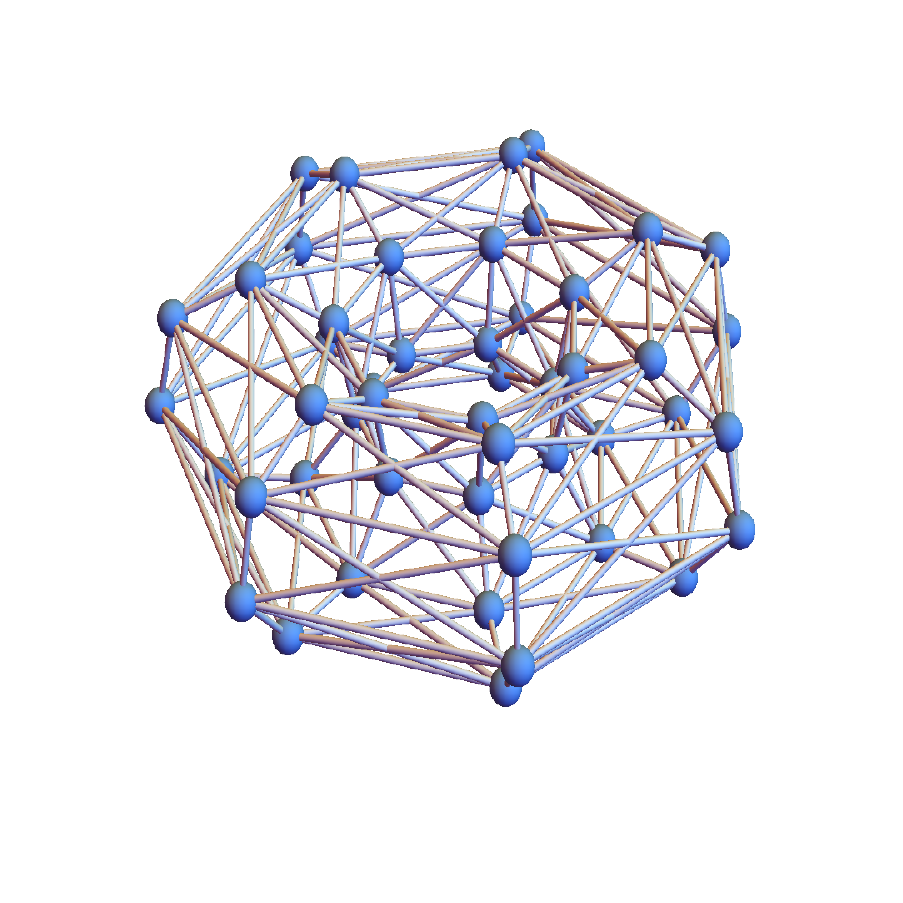}}
\label{King}
\caption{
The product graph $L_7 \cdot L_7$, where $L_7$ is the linear graph is
the king connection graph on a $7 \times 7$ board.
The product graph $C_7 \cdot C_7$, where $C_7$ is the cycle graph is
the king connection graph on a toroidal $7 \times 7$ board. 
}
\end{figure}

\paragraph{}
The Shannon product is an associative product. Together with addition as disjoint union,
one has now a {\bf number system} in which the graphs are the numbers.

\begin{figure}[!htpb]
\scalebox{0.5}{\includegraphics{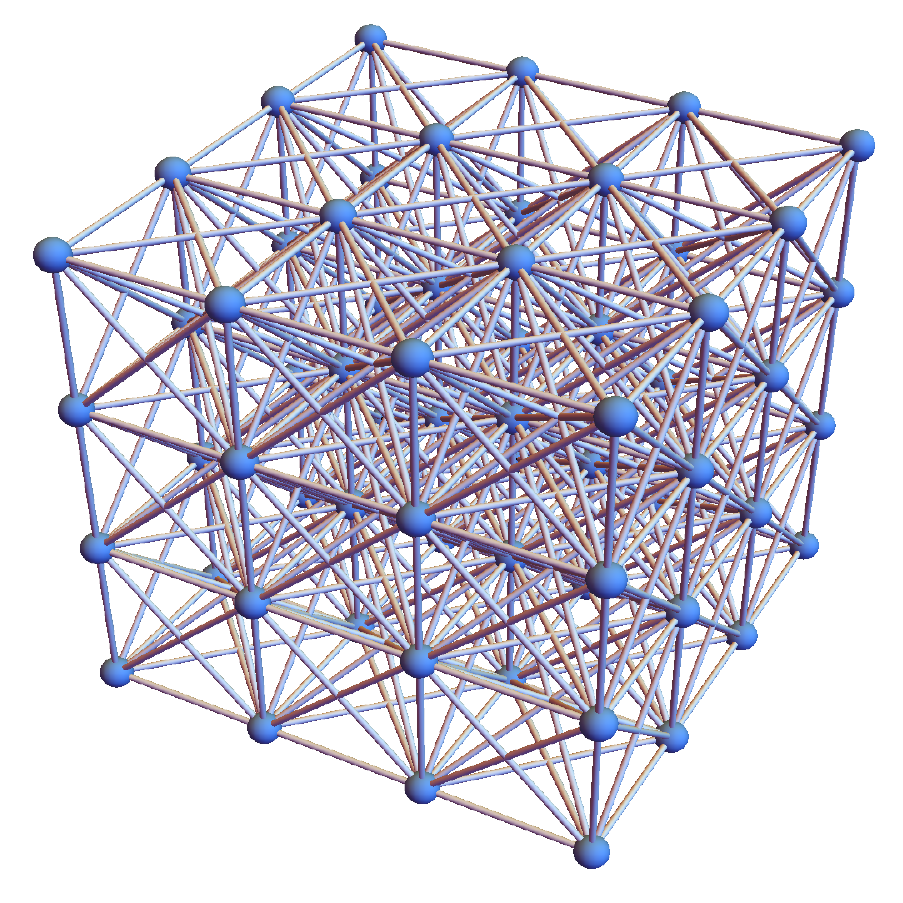}}
\scalebox{0.5}{\includegraphics{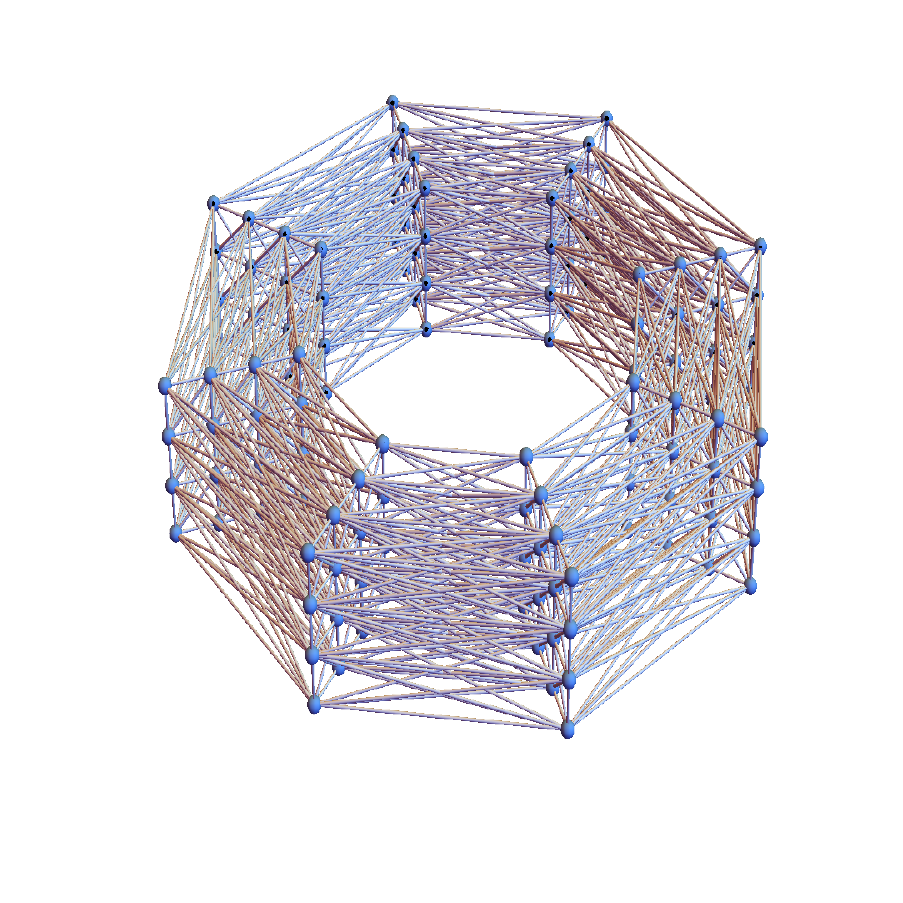}}
\label{King}
\caption{
The $3$ dimensional cube $L_4 \cdot L_4 \cdot L_4$ and the solid
2-torus $C_4 \cdot C_4 \cdot L_8$ are all products of 
three graphs. We have $(L_4 \cdot L_4) \cdot C_8 = L_4 \cdot (L_4 \cdot C_8)$. 
}
\end{figure}

\paragraph{}
Let us denote a graph with $G$ and call $\alpha(G)$  the independence number of the graph. 
This is the maximal number of points one can chose such that none are connected to each other.
On the cyclic graph with $2n$ elements, one has $\alpha(C_{2n})=n$. 
On a cyclic graph with $2n+1$ elements, one has also $\alpha(C_{2n+1})=n$. 
Now lets look at the product. We have seen $\alpha(C_4 \cdot C_4) = 4$ and 
$\alpha(C_5 \times C_5) = 5$. The capacity of the two point channel for $C_4$ is now 
$\alpha(C_4 \cdot C_4)^{1/2} = 2$ which agrees with $\alpha(C_4)$. The capacity of the
two point channel for $C_5$ is $\alpha(C_5 \times C_5)^{1/2} = \sqrt{5}$. This is better than 2.
Since this went so well, we do better taking more copies of the channel? 

\paragraph{}
The Shannon capacity $\Theta(G)$ is the limit $\alpha(G^n)^{1/n}$ for $n$ going to infinity.
The cyclic graph with 5 elements $C_5$ was the first graph, where Shannon could not compute
the capacity yet. All other graphs with $5$ or less vertices, he could handle. In the case 
$n=3$ we deal with 3D toral chess. How many kings can we place?
It might surprise, but it is actually a bit worse. We can place $11$ kings. Now $11^{1/3}$ is $2.2239$
which is smaller than $\sqrt{5}=2.236$. How many can we place on a $3$-torus?

\begin{figure}[!htpb]
\scalebox{0.8}{\includegraphics{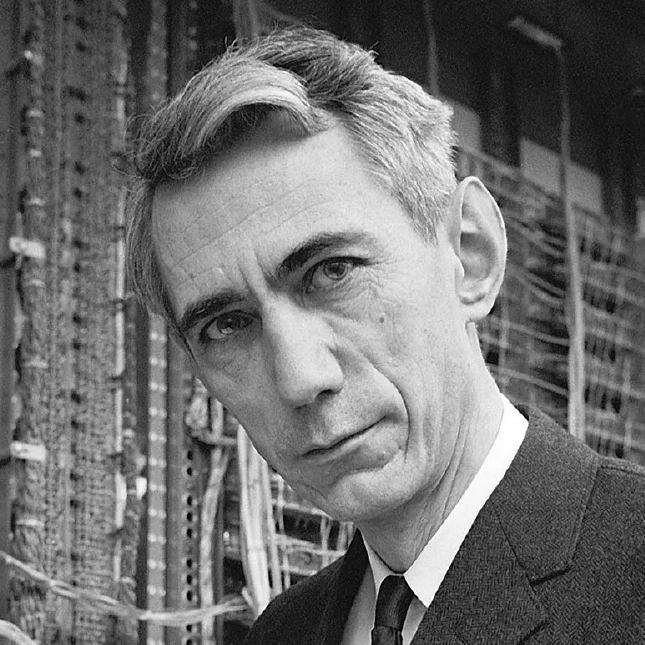}}
\scalebox{0.8}{\includegraphics{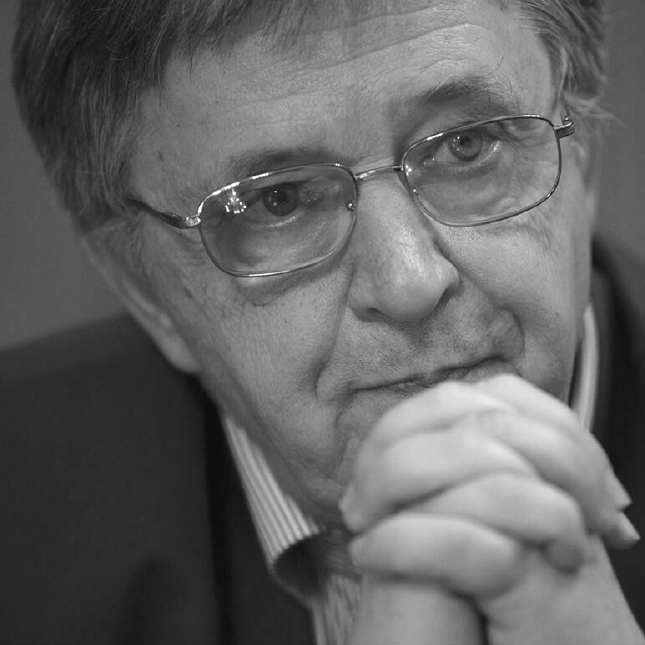}}
\label{King}
\caption{
Only 20 years after Shannon it became possible to find $\Theta(C_5)$. 
The figure shows Claude Shannon left and Laslo Lovasz to the right. 
}
\end{figure}

\section{Entanglement}

\paragraph{}
let us attach a quantum mechanical systems to each vertex $x$ in the graph.
Since vectors of unit length are called {\bf states}, an umbrella $U$ is the process of
attaching a ``state" to every node. 
If $U$ is an umbrella for $G$ and $V$ is an umbrella for $H$
then $UV(x,y) = U(x) V(y)$, the {\bf tensor product} is an umbrella for $G \cdot H$.
Indeed, if $(a,b)$ and $(c,d)$ are not connected, then $UV(a,b)$ and 
$UV(c,d)$ are still perpendicular.  The tensor product of two fields is
an ``entangled state" in quantum mechanics. 

\paragraph{}
The assumption that states of disconnected parts are perpendicular is natural
as there can not be any direct interaction. Any possible dynamics would come through quantum
fields, which means to look  look at powers of the geometry. If we look at the umbrella stick
as the vacuum, an optimal umbrella maximizes the minimal correlation $U(x) \cdot c$. 
The Lovasz number is compatible with multiplication. 

%Example: $U(x)$ for $G$ and $V(x)$ for $H$ with maximal $1/(c \cdot U(x))^2$
%rsp $1/(d \cdot V(x))^2$. Now we produce the umbrella $(U,V)$ with vacuum $(c/\sqrt{2},d/\sqrt{2})$.
%Now $((c,d) \cdot (U,V) ) = ((c.U) + (d.V))^2

\section{Pure states}

\paragraph{}
Here is an observation which we find interesting despite its simplicity. Instead of unit vectors, we can 
look at {\bf density matrices}, non-negative definite symmetric matrices of trace $1$ and attach such 
a density matrix at each node of the graph. The {\bf Hilbert-Schmidt inner product} ${\rm tr}(A^T B)$ 
allows to define the Lovasz number in the same way. 
With a density matrix $A(x)$ attached to each vertex of a graph $G$, we 
ask for $A(x) \cdot A(y) = {\rm tr}(A(x)^T A(y))=0$ for non-adjacent vertices. Also the {\bf vacuum state}
$c$ can be a density matrix. We still have $\alpha(G) \leq \theta_{A,c}(G) = {\rm max}_x (c \cdot c)/(c \cdot A(x))^2$ 
and because $|A(x)|^2 = \sum_x \lambda_x^2 \leq \sum_x \lambda_x = 1$, the density matrices have norm 
bounded by $1$ and the norm is $1$ if and only if we have a pure state. It follows that an optimal
Lovasz umbrella attaches a pure state at each vertex. The Lovasz number does not change when making an evolution
of an isolated system, where pure states are evolved. If we look at a quantum evolution of a 
multi-particle system however, the Lovasz number gets bigger in general because entangled umbrellas
are less optimal. 

\section{A difficult problem}

\paragraph{}
The definition of Shannon capacity involves the independence number $\alpha$. This is a difficult
number to compute. It actually is what one knows to be an NP complete problem. The reason is 
that the graph complement $\overline{G}$ which is the graph in which edges and non-edges are switched
has a clique for each independent set. A clique is also called a facet, a maximal simplex in the 
Whitney complex. 
Formally, this means $\alpha(G) = \omega(\overline{G})$, where
$\omega(G)$ is the clique number the number of vertices in a maximal clique. NP complete means that
if there was a way to find the independence number fast, in a time which is polynomial in the
number of vertices of the graph, then all NP problems can be computed fast. The famous P versus NP problem,
one of the millenium problems asks to prove that the class P is different from NP. 

\section{The heptagon}

\paragraph{}
One can get lower bounds of the Shannon entropy by computing the independence number 
$\alpha(G^n)^{1/n} $ for some $n$. For $G=C_7$ and $n=2$, we get $\sqrt{\alpha(G^2)} = \sqrt{10}$
because we can place $10$ kings on a toroidal $7 \times 7$ chess board which are not neighbors.
Now, for a $7 \times 7 \times 7$ chess board, this implies that at least $30$ 
kings can be placed. We have from $d=2$ already the lower bound $10^{1/2} = 3.16228$. 
An upper bound for odd $C_n$ with odd $n$ are given with 
$\theta(C_n,U) = n \cos(\pi/n)/(1+\cos(\pi/n))$ which is $\sqrt{5}$ for $n=5$ and
$(7/2) \cos(Pi/7)/\cos^2(\pi/14) = 3.31767\dots$ for $n=7$. \cite{Hales1973,KnuthSandwich}
%  f[n_]:=n Cos[Pi/n]/(1+Cos[Pi/n])

\begin{center}
The Shannon capacity of $C_7$ is between $3.16228$ and $3.31767$. 
\end{center}

\paragraph{}
After Shannon, it was \cite{Hales1973} who in his doctoral thesis
at Harvard under the guidance of Andrew Gleason wrote about strong products of graphs
\cite{Hales1973}. There, the {\bf Rosenfeld number} $\rho(G) = \sup_{v} f(v)$ among all 
functions $f:V \to [0,\infty)$ with $\sum_{v \in x} f(v) \leq 1$ for every simplex $x$. 
The clique covering number $\sigma$ (introduced by Shannon) is the number of cliques in a minimum
clique cover of $G$.  Then $\alpha(G) \leq \rho(G) \leq \sigma(G)$. For $G=C_{2k+1}$ for
example, one has $\alpha(G)=k, \rho(G)=k+1/2, \sigma(G)=k+1$. 
The number $\rho$ is multiplicative. If $G$ contains no $C_{2k+1}$, then 
$\Theta(G)=\lambda(G)=\alpha(G)$. This shows how important it is to study odd cycles.

\section{Toroidal 3D chess}

\paragraph{}
In order to get a better lower bounds on the Shannon capacity
in the $C_7$ case, we would need to place 32 kings.

\begin{figure}[!htpb]
\scalebox{0.7}{\includegraphics{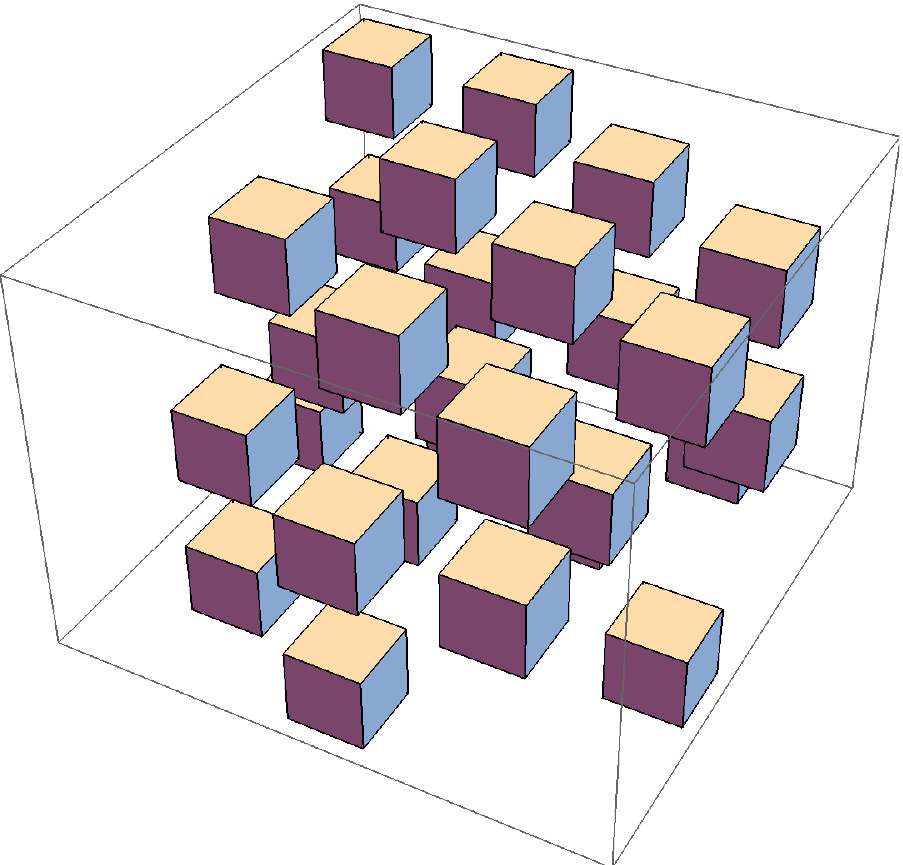}}
\label{10King}
\caption{
How many kings can be placed on a $7 \times 7 \times 7$ toral
chess board? We see here a solution with 30 kings. This was just
obtained by placing 10 king solutions in layer $1,3$ and $5$. We
believe that this can be improved. Going through all the
cases with backtracking is too expensive.
}
\end{figure}

\section{Cube chess}

\paragraph{}
There are already many variants of chess known.  There is a circular 3-people chess 
for example. In higher dimensions, there is ``space chess", ``Raum chess", ``Strato chess"
and a chess appearing in Star Trek. Let us add an other variant which we have not seen
anywhere yet. It is a $8^3$ ``cube chess", played on a $8 \times 8 \times 8$ cube. 
The rules agree all with the usual chess when restricted to two dimensions but of course
we have to make adaptations: each of the two kings is surrounded by queen and two princesses 
(with the same power than the their mother). The royal core family of 4 people
is surrounded by body guards (12 bishops) which can move along any diagonal including
space diagonals, then comes the cavalry (20 knights), surrounded
finally by artillery (28 rooks). There are 4+12+20+28=64 main figures of each color. 
As usual, a queen combines the abilities of the bishops and rooks and the
king can move to a neighboring field. The positions in the second
and second last plane are covered with 64 soldiers (pawns), which can move 
in the third direction, capture diagonally including on passant, become a piece of 
choice when reaching the other end. The king can castle with any of the 
rooks in the same row or column as long as the space between them is free.
This $8 \times 8 \times 8$ cube chess is a very direct dimensional adaptation of 
our square chess. A first question of course is whether there is an obvious  
a winning strategy. 

\paragraph{}
One could look at smaller versions. 
A $4^3$ cube chess with all fields covered probably can be completely analyzed.
On each side, there are 4 rooks at the border, 8 bishops at the boundary
king, queen, and two horses are in the center. All paws are placed on the
second or third layer. This $4^3$ chess leads to a bloody battle with 
lots of initial losses for the second player so that very likely there is a winning
strategy for the first player. 

\begin{figure}[!htpb]
\scalebox{0.06}{\includegraphics{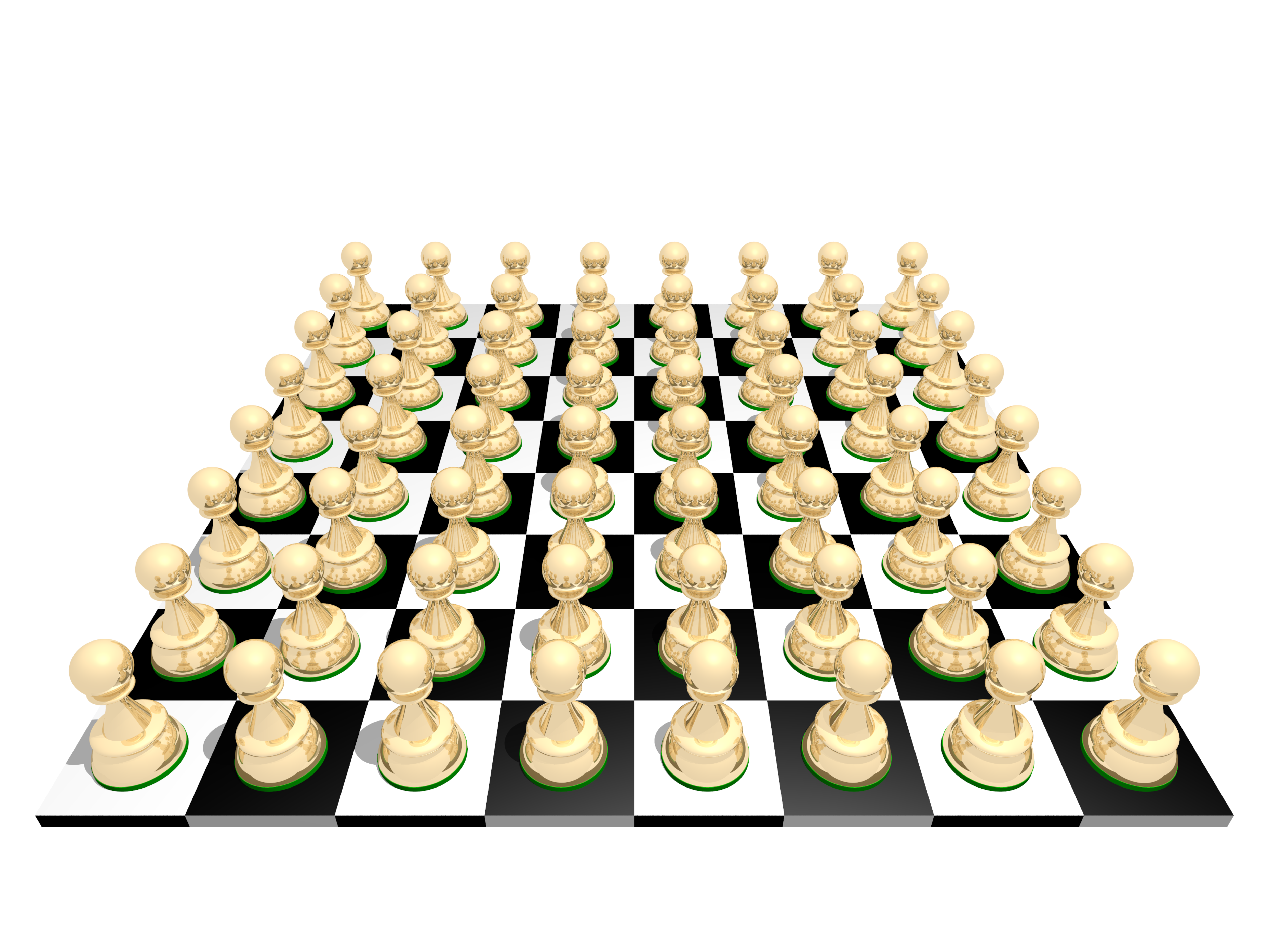}}
\scalebox{0.06}{\includegraphics{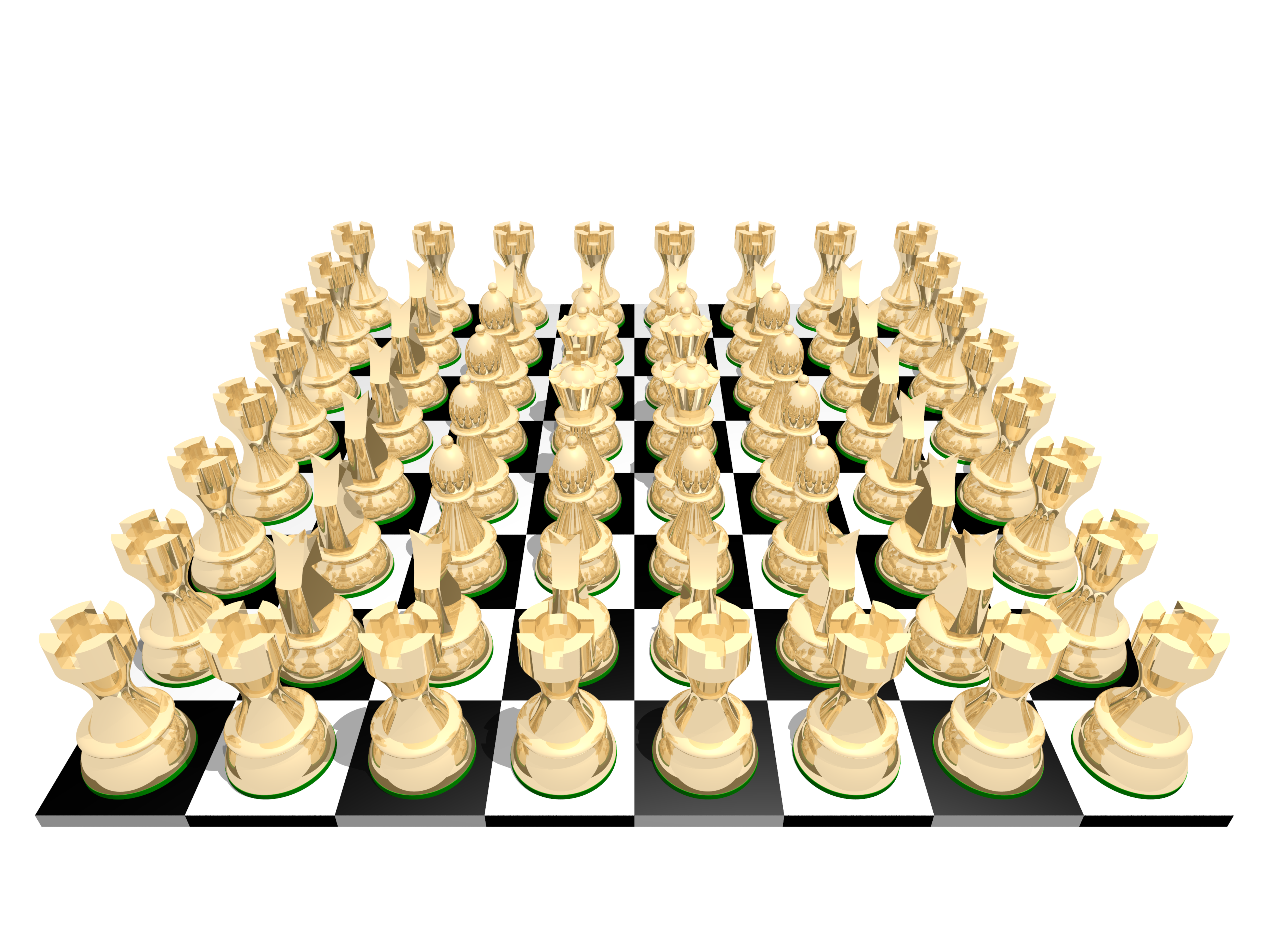}}
\label{Cube Chess}
\caption{
The first and second layer of the white side of cube chess
consists of 1 king, 3 queens, 12 bishops, 20 knights and 28 rooks
and 64 pawns.
}
\end{figure}

\section{More graphs}

\paragraph{}
There are other graphs for which we can compute the Shannon capacity. 
Shannon computed the capacity of all graphs $\leq 4$. 
For connection graphs, the capacity is the number
of zero dimensional elements in $G$ \cite{ComplexesGraphsProductsShannonCapacity}.
For Barycentric graphs the capacity is
bound from below also by the number of facets in $G$, elements in $G$ 
which are not contained in a larger set and the number of $0$-dimensional elements.
For $G=K_2$ we have $\Theta(G_1) \geq 3$, the number of vertices of $G_1$
but for $G$ producing the figure $8$ graph, we have $\Theta(G_1)$ is the number of 
edges of $G_1$ which is larger than the number of vertices. 

\begin{figure}[!htpb]
\scalebox{0.25}{\includegraphics{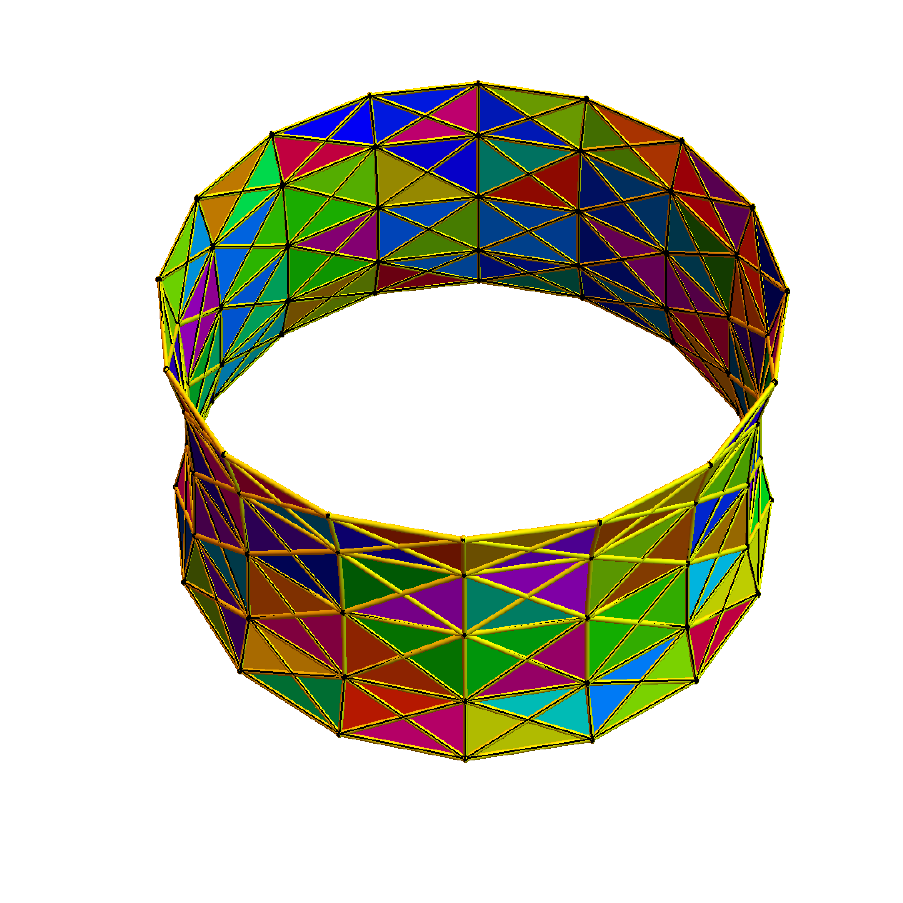}}
\scalebox{0.25}{\includegraphics{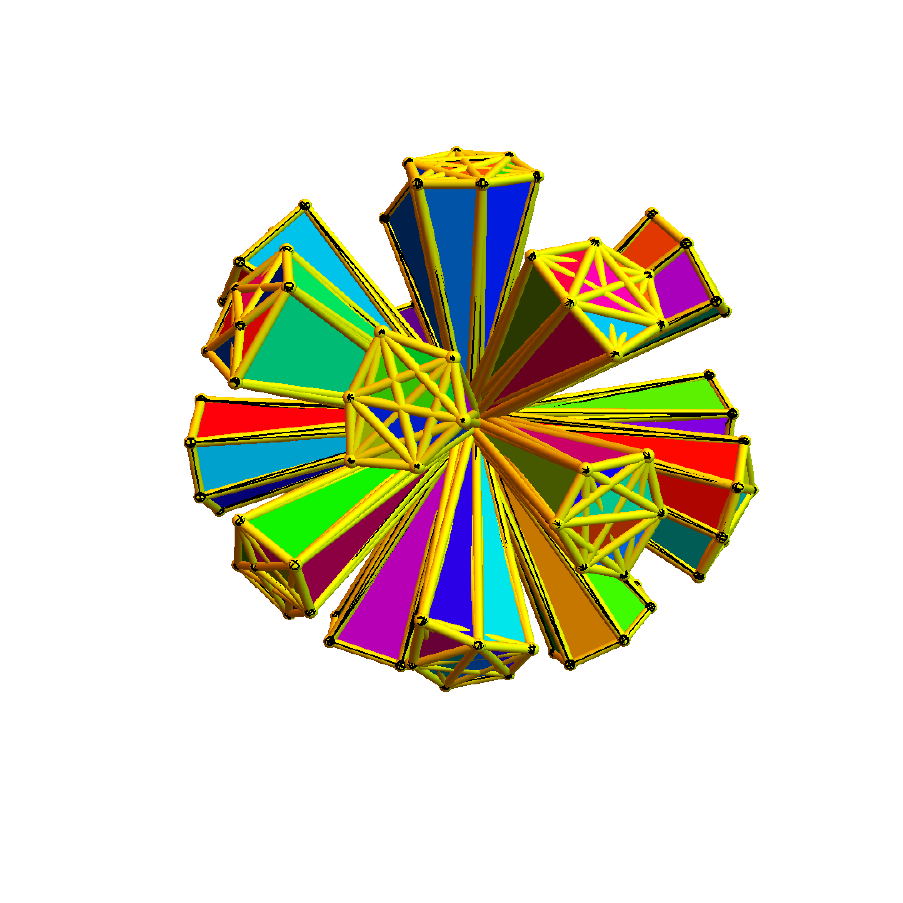}}
\scalebox{0.25}{\includegraphics{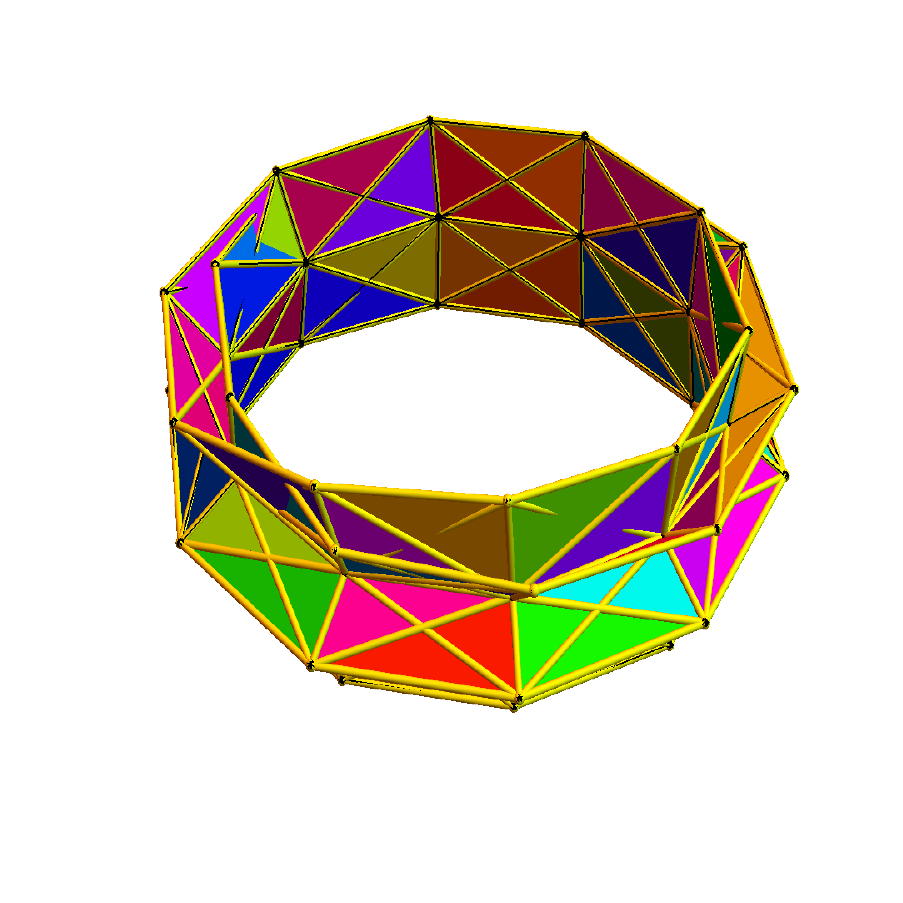}}
\scalebox{0.25}{\includegraphics{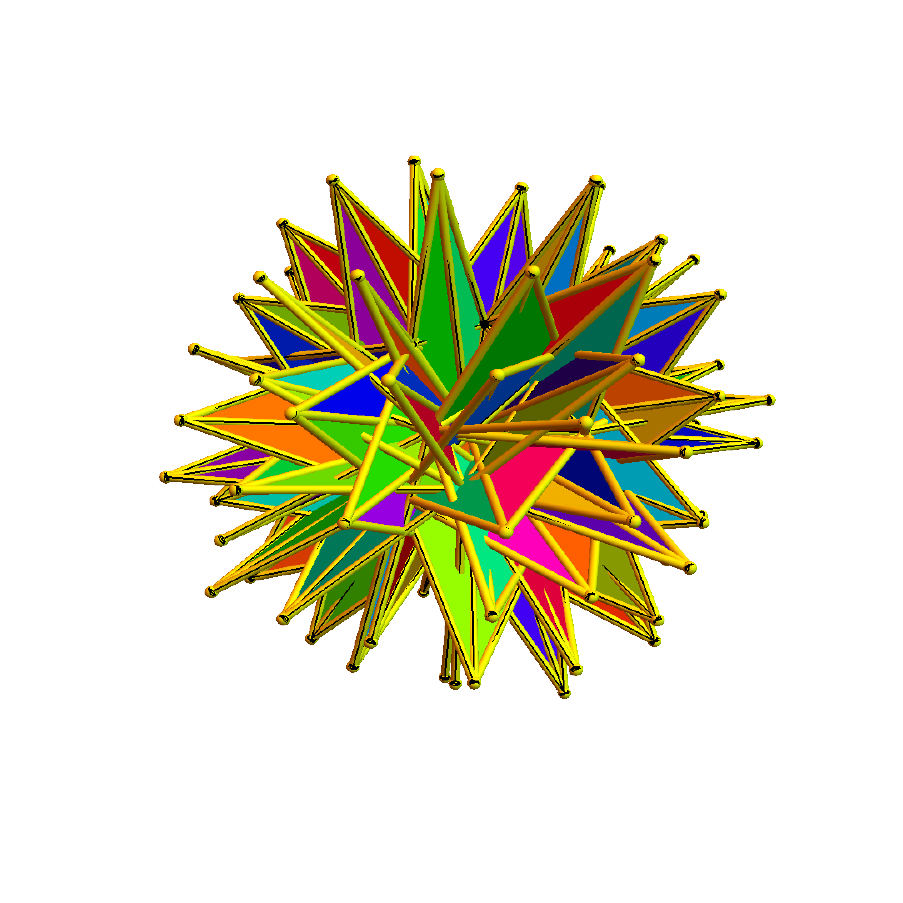}}
\scalebox{0.25}{\includegraphics{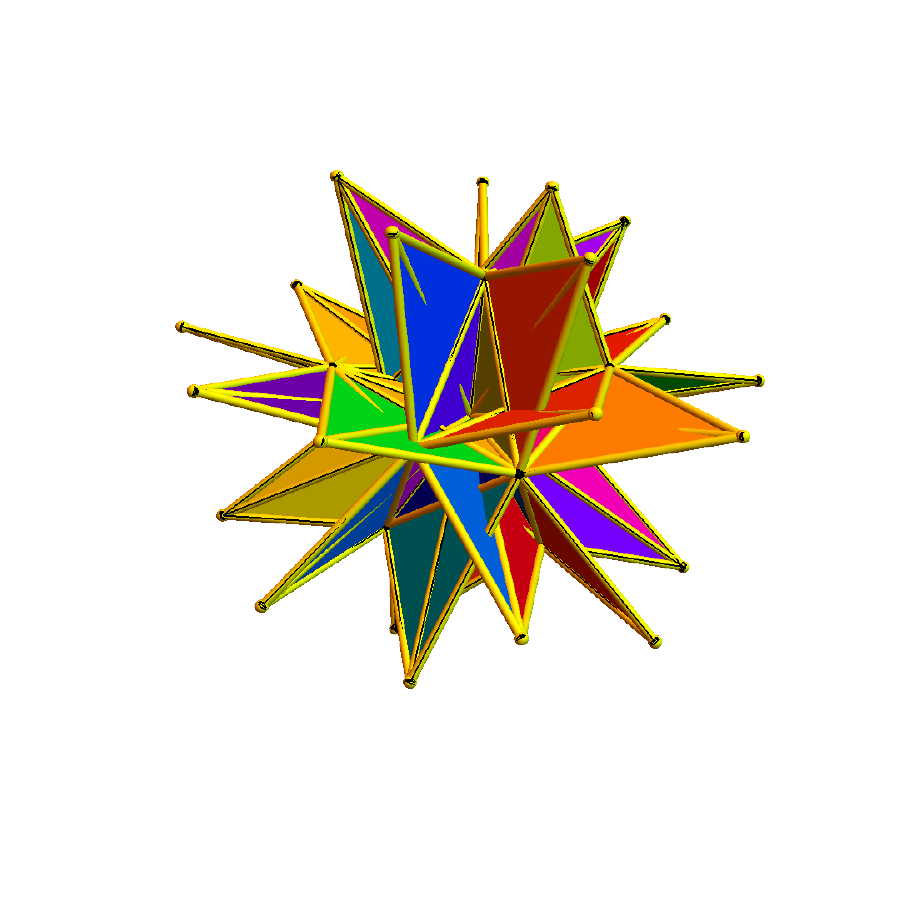}}
\scalebox{0.25}{\includegraphics{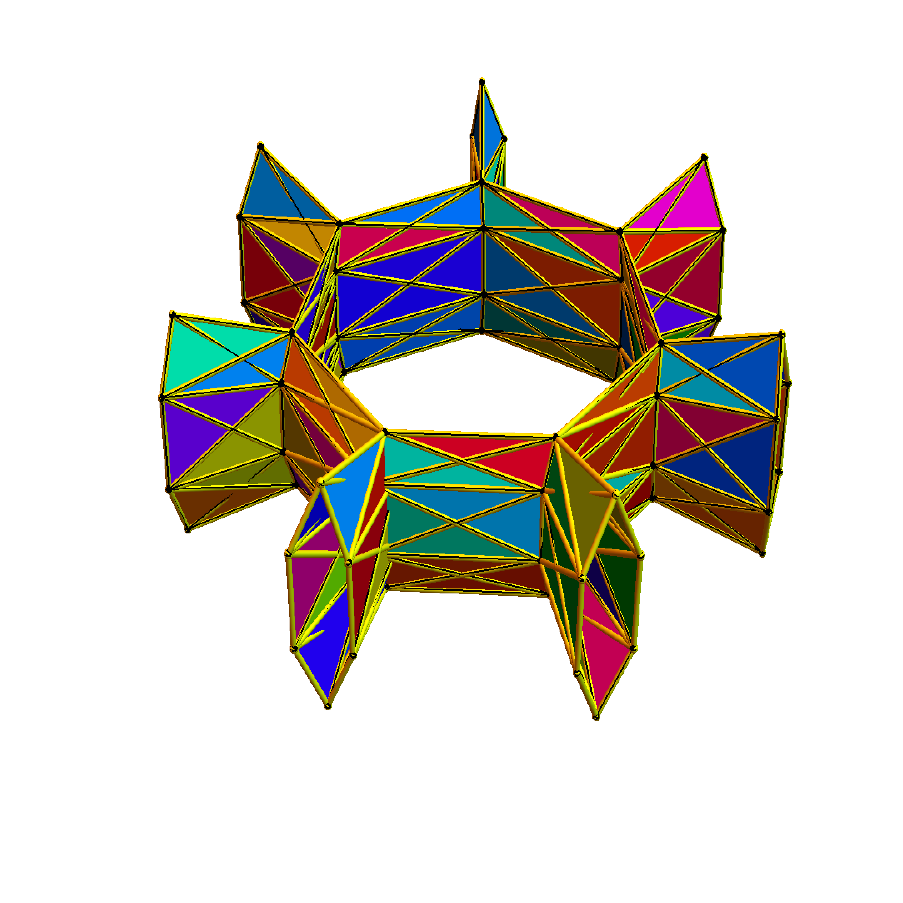}}
\scalebox{0.25}{\includegraphics{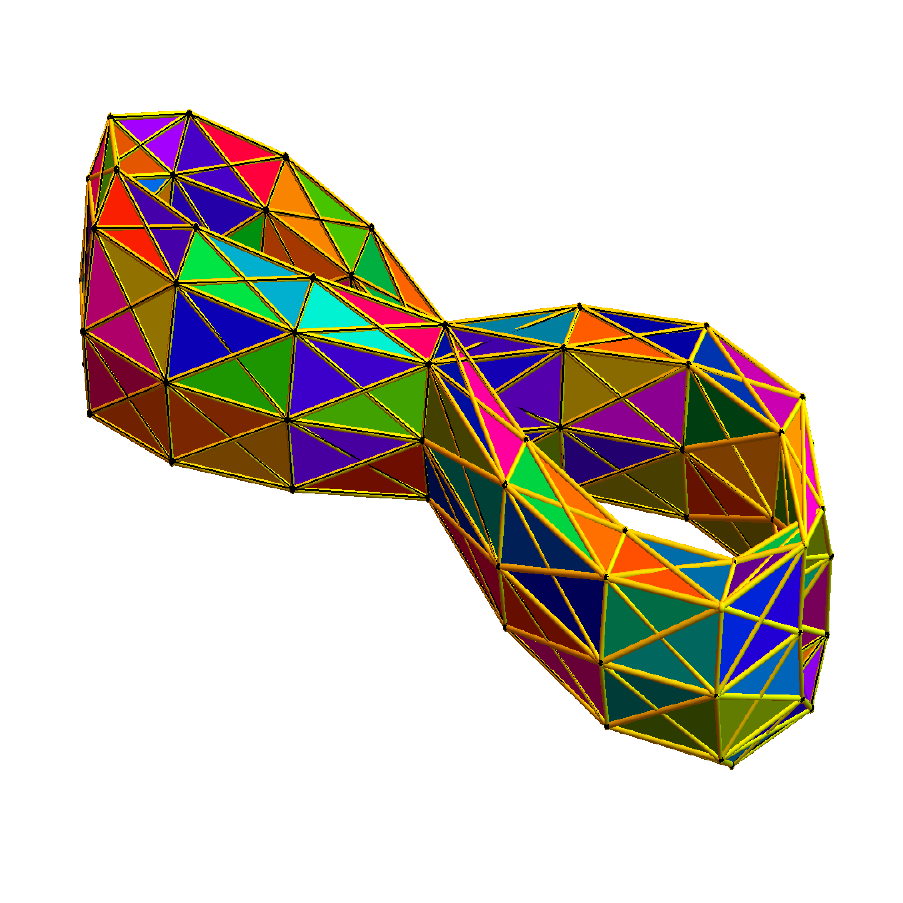}}
\scalebox{0.25}{\includegraphics{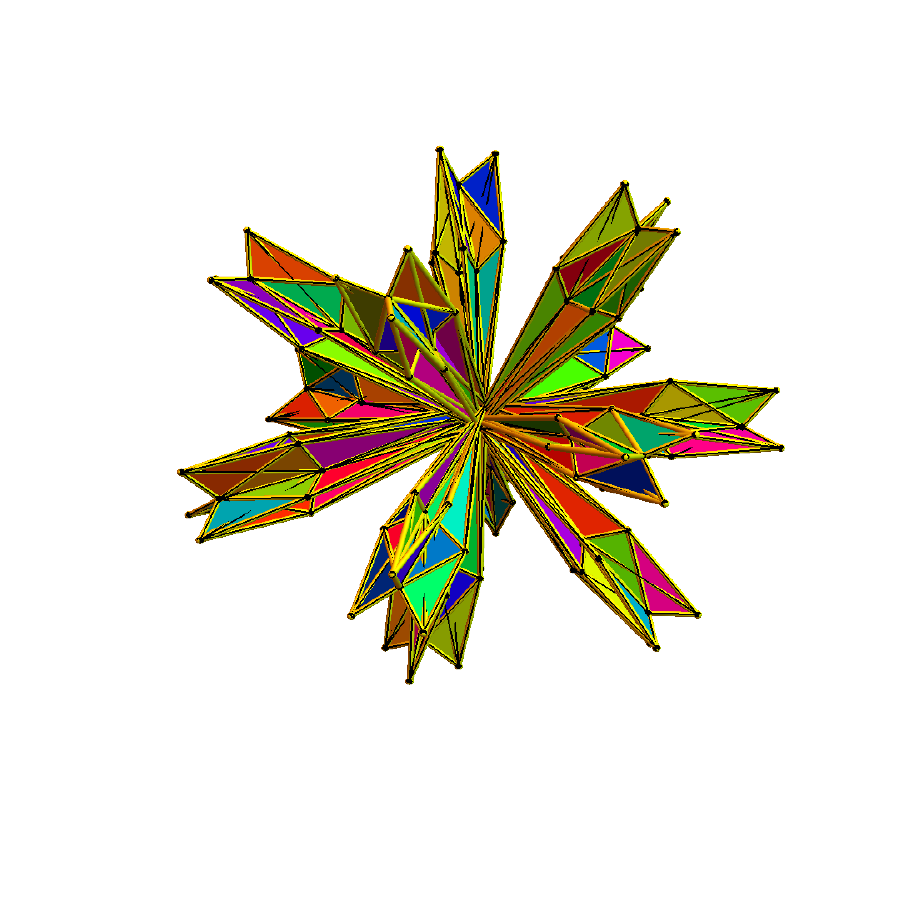}}
\scalebox{0.25}{\includegraphics{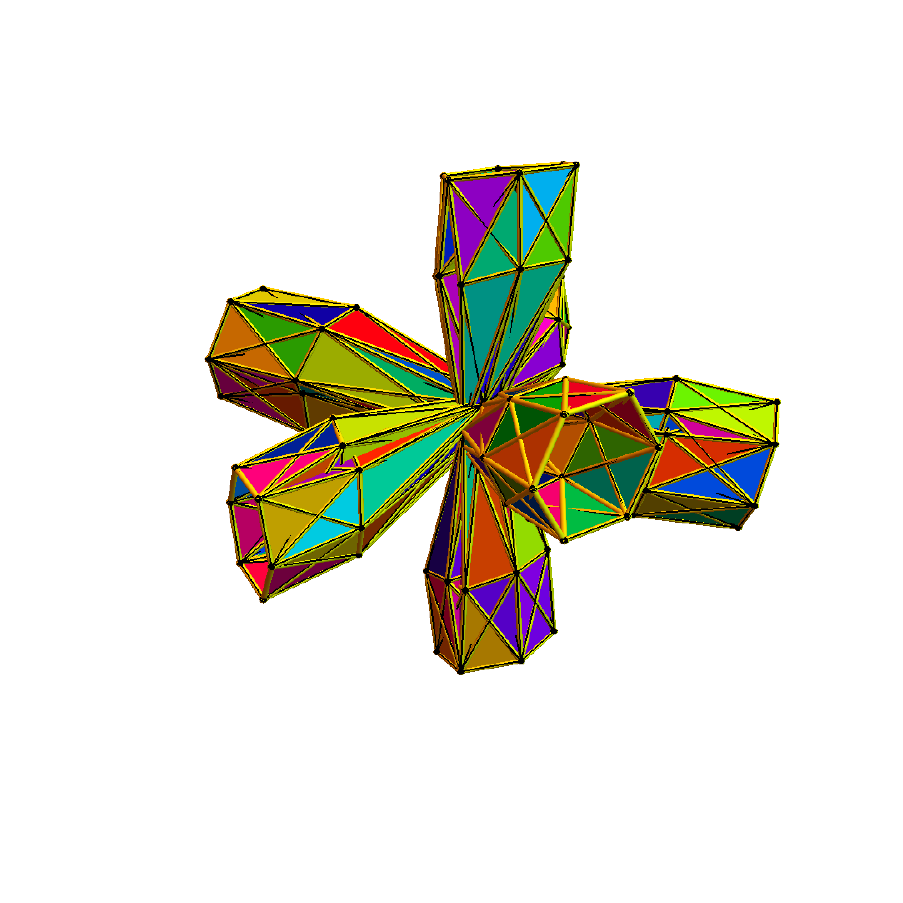}}
\scalebox{0.25}{\includegraphics{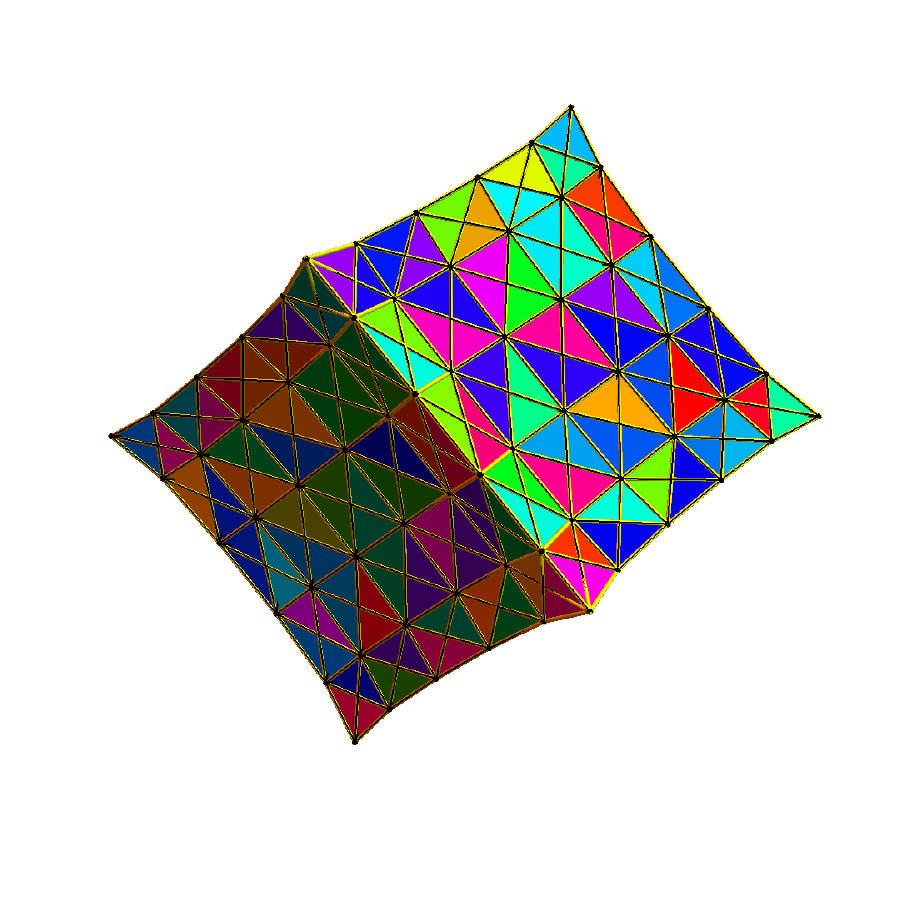}}
\scalebox{0.25}{\includegraphics{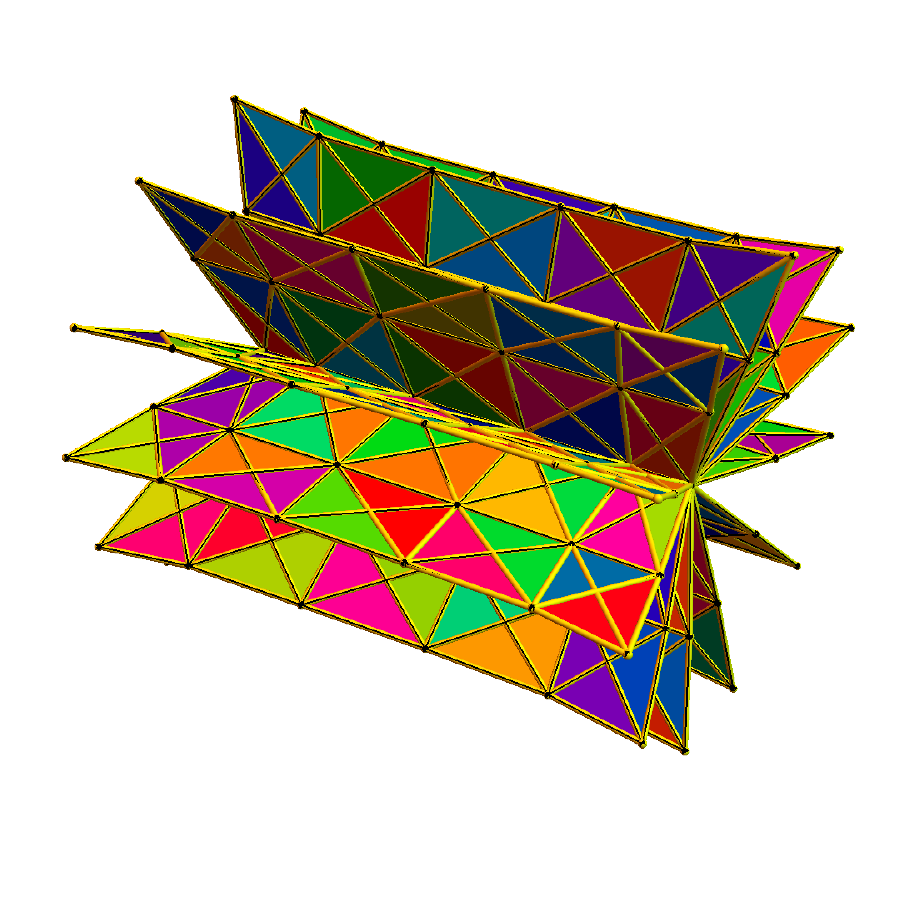}}
\scalebox{0.25}{\includegraphics{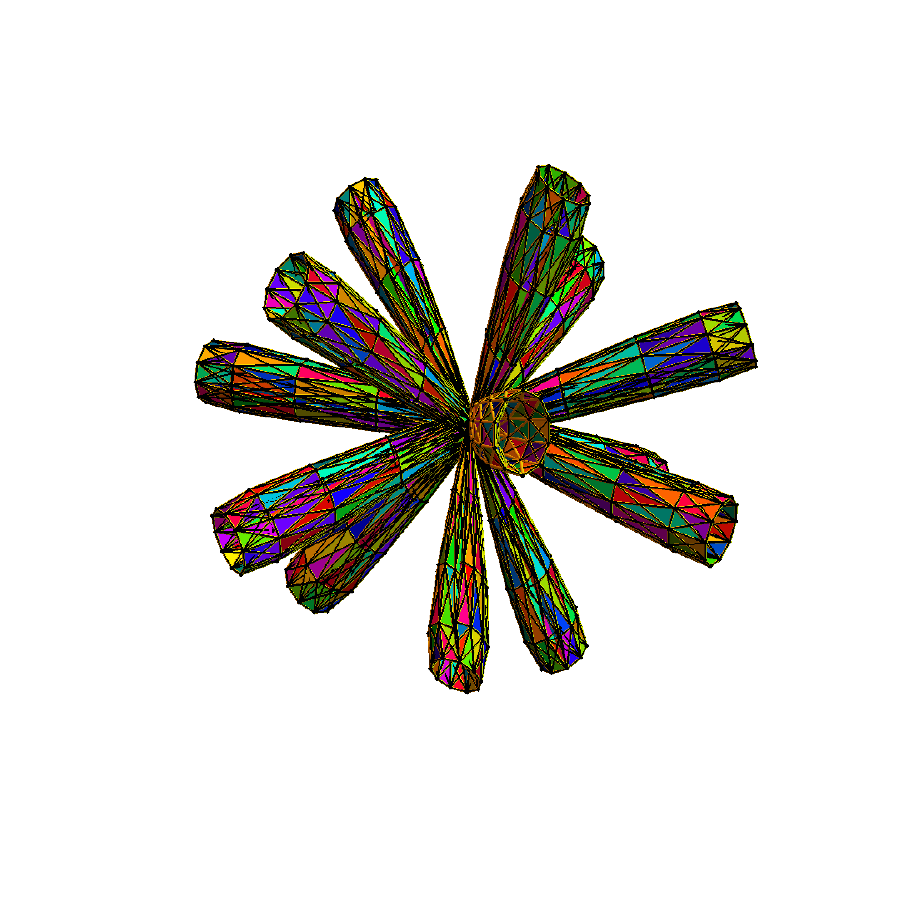}}
\scalebox{0.25}{\includegraphics{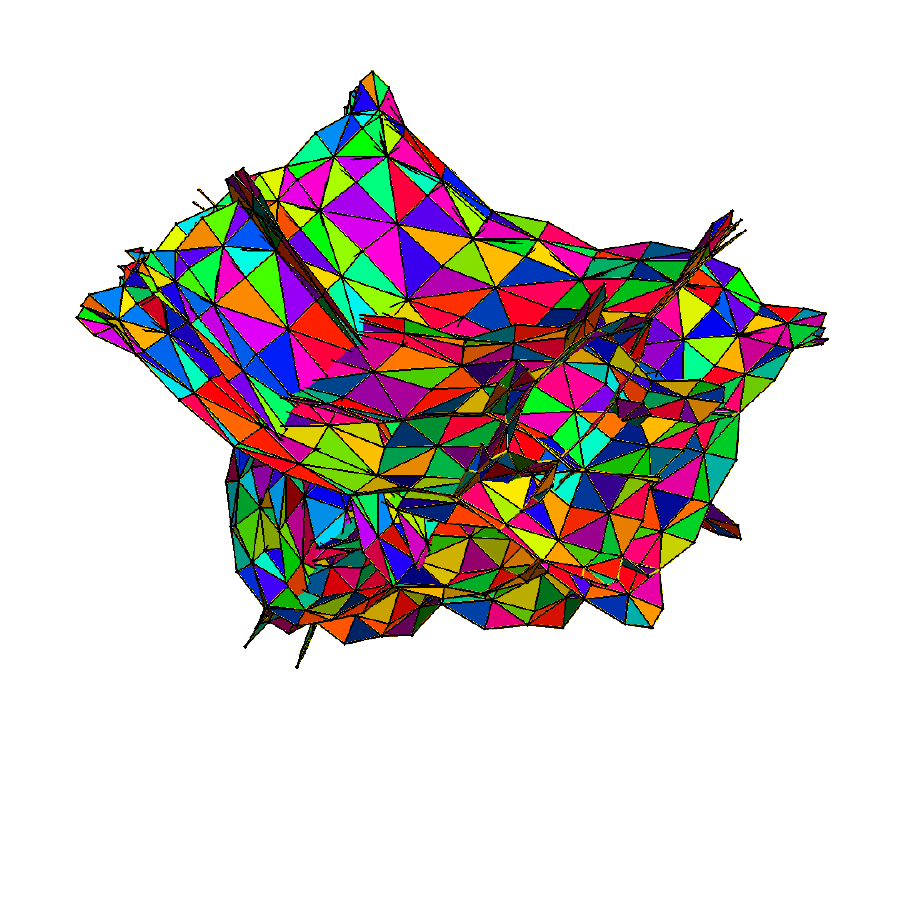}}
\scalebox{0.25}{\includegraphics{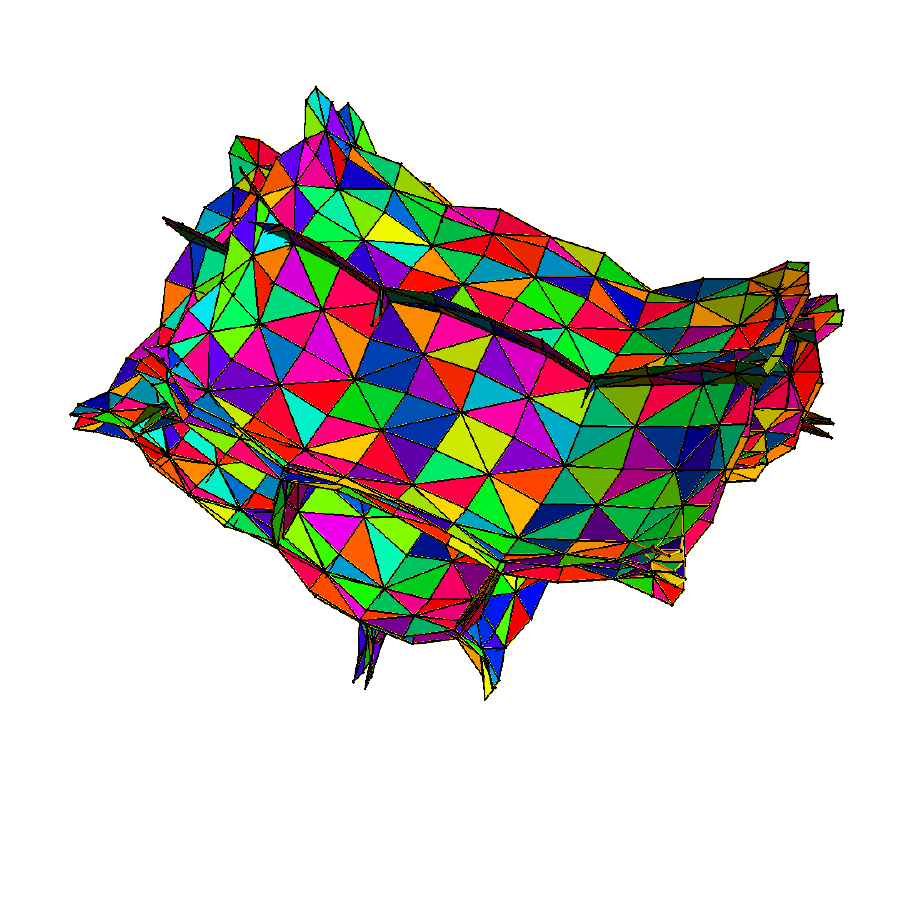}}
\scalebox{0.25}{\includegraphics{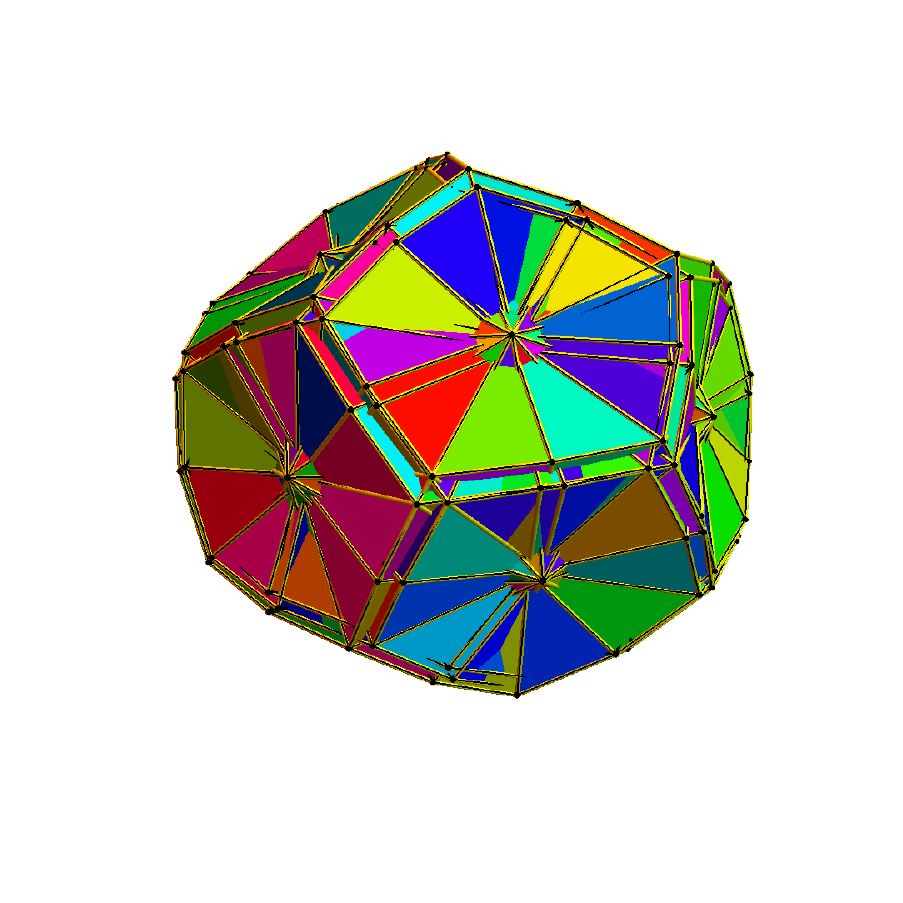}}
\scalebox{0.25}{\includegraphics{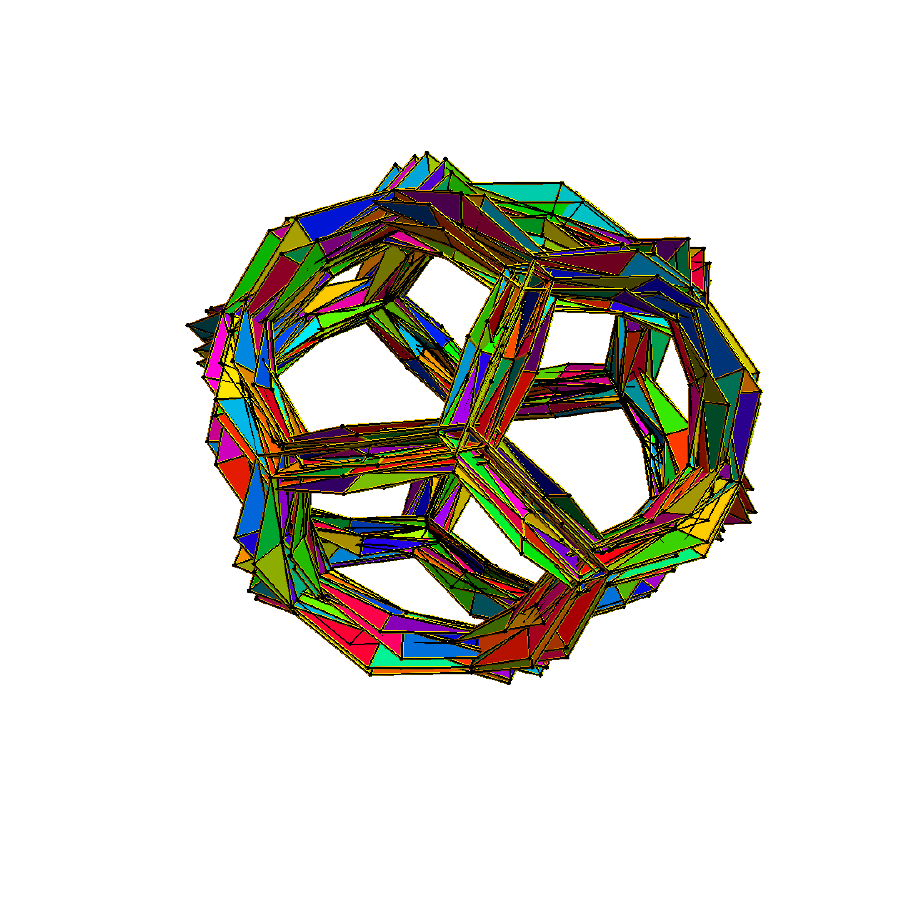}}

\label{Products}
\caption{
Examples of graph products. 
}
\end{figure}

\section{Remarks}

\paragraph{}
One can try to use the {\bf dual picture} for the computation of capacity. 
The clique number $c(G)$ of a graph is the independence number of the graph 
complement so that the Shannon capacity measures the exponential growth rate
of the clique number of Sabidussy products, which is the product dual to the Shannon product.  \\

Here is the {\bf sandwich theorem} and its dual version: 
$$ \alpha(G) \leq \theta(G) \leq \sigma(G)   $$
$$ \omega(G) \leq \theta(\overline{G}) \leq \xi(G) $$
where $\alpha(G)$ is the {\bf independence number}, $\theta(G)$ is the {\bf Lovasz number},
$\sigma(G)$ the {\bf clique covering number} (the minimal number of vertex disjoint cliques), 
$\omega(G)$ the {\bf clique number} (the size of the largest clique) 
and $\xi(G)$ the {\bf chromatic number}. \\

Since the clique covering number $\sigma(G)$, an upper bound for $\Theta(G)$,
is the chromatic number of the graph complement $\xi(\overline{G})$. 
So, the growth of chromatic numbers under large products matters. Example:
for the $n$-sphere $P_2 \oplus P_2 \oplus P_2 \oplus P_2$ is dual to $4 K_2$. 
The dual of the star graph is $1+K_n$. 

\paragraph{}
Graph complements are topologically rich already in simple cases.
Lets look at cycle graphs. The dimension grows like $n/2$ but cohomology only grows
like $n/3$. For $C_3$ we have the three point graph $P_3$, for $C_4$ we have $K_2 + K_2$
the only disconnected case, the pentagon $C_5$ is self-dual $\overline{C_5}=C_5$.
The dual of $C_7$ is a discrete M\"obius strip.

\paragraph{}
Graph theory has also a differential geometric component. There is a curvature $K(x)$ which 
adds up to the {\bf Euler characteristic}. 
Using index expectation we have shown that the curvature of the Shannon product of two graphs is the tensor
product of the curvatures of the individual graphs \cite{GraphProducts}. 
We also studied recently the curvature of graph complements \cite{GraphComplements} of cyclic or path graphs.
The curvature of path graphs converges universally to an attractor.  

\paragraph{}
We have once studied (mostly experimentally) a couple of variational problems in \cite{KnillFunctional}.
For any interesting quantity on graphs, one can ask whether there is a Gauss-Bonnet formula
meaning that the quantity can be computed by summing up a {\bf local quantity}. 
For independence numbers or chromatic numbers for example, there exist no local quantity for a Gauss-Bonnet
theorem. The reason is that
both quantities can depend on global properties like parity. We can have pairs of graphs $G,H$
where all unit spheres are isomorphic but where $G,H$ have different chromatic number.

\bibliographystyle{plain}

\begin{thebibliography}{10}

\bibitem{Quadruplehelix}
M.~Antonio, A.~Ponjavic, and et~al. A.~Radzeviius.
\newblock Single-molecule visualization of {DNA G-quadruplex} formation in live
  cells.
\newblock {\em Nat. Chem.}, 12:832--837, 2020.

\bibitem{Bohman2003}
T.~Bohman.
\newblock A limit theorem for the {Shannon} capacities of odd cycles i.
\newblock {\em Proc. Amer. Math. Soc}, 131:3559--3569, 2003.

\bibitem{Gquartet}
A.~Calzolari, R.~Di Felice, and E.~Molinari.
\newblock G-quartet biomolecular nanowires.
\newblock {\em Appl. Phys. Lett.}, 80:3331--3333, 2002.
\newblock https://doi.org/10.1063/1.1476700.

\bibitem{Haemers1978}
W.~Haemers.
\newblock An upper bound for the {Shannon Capacity} of a graph.
\newblock {\em Colloquia Mathematica Societatis Janos Bolyai}, 25:267--272,
  1978.

\bibitem{Hales1973}
R.S. Hales.
\newblock Numerical invariants and the strong product of graphs.
\newblock {\em J. of Combinatorial Theory}, 15:146--155, 1973.

\bibitem{KnillFunctional}
O.~Knill.
\newblock Characteristic length and clustering.
\newblock {{\\}http://arxiv.org/abs/1410.3173}, 2014.

\bibitem{ComplexesGraphsProductsShannonCapacity}
O.~Knill.
\newblock {Complexes, Graphs, Homotopy, Products and {Shannon} Capacity}.
\newblock {{\\}https://arxiv.org/abs/2012.07247}, 2020.

\bibitem{GraphProducts}
O.~Knill.
\newblock The curvature of graph products.
\newblock https://arxiv.org/abs/2107.08563, 2021.

\bibitem{GraphComplements}
O.~Knill.
\newblock Graph complements of circular graphs.
\newblock https://arxiv.org/abs/2101.06873, 2021.

\bibitem{KnuthSandwich}
D.~Knuth.
\newblock The sandwich theorem.
\newblock {\em Electronic journal of Combinatorics}, 1, 1994.

\bibitem{Lovasz1979}
L.~Lovasz.
\newblock On the {S}hannon capacity of a graph.
\newblock {\em IEEE Transactions on Information Theory}, 25:1--7, 1979.

\bibitem{Matousek}
J.~Matousek.
\newblock {\em Thirty-three Miniatures}, volume~53 of {\em Student Mathematical
  Library}.
\newblock AMS, 2010.

\bibitem{PritchardChess}
D.B. Pritchard.
\newblock {\em The Classified Encyclopedia of Chess Variants}.
\newblock John Beasley, 2007.
\newblock originally printed Biddles Ltd, Kings's Lynn.

\bibitem{Rosenfeld1967}
M.~Rosenfeld.
\newblock On a problem of {C. E. Shannon} in graph theory.
\newblock {\em Proc. Amer. Math. Soc.}, 18:315--319, 1967.

\bibitem{Sabidussi}
G.~Sabidussi.
\newblock Graph multiplication.
\newblock {\em Math. Z.}, 72:446--457, 1959/1960.

\bibitem{Shannon1956}
C.~Shannon.
\newblock The zero error capacity of a noisy channel.
\newblock {\em IRE Transactions on Information Theory}, 2:8--19, 1956.

\bibitem{MindAtPlay}
J.~Soni and R.~Goodman.
\newblock {\em A mind at play}.
\newblock Simon and Schuster, 2017.

\end{thebibliography}

\end{document}